\documentclass{article}

\usepackage{arxiv}

\usepackage[utf8]{inputenc} 

\usepackage{hyphenat}
\usepackage{multirow}
\usepackage{balance} 

\usepackage[numbers]{natbib}
\usepackage{algorithmic}
\usepackage{algorithm}
\usepackage{amsthm}
\usepackage{amssymb}
\usepackage[export]{adjustbox} 
\usepackage{amsmath}
\usepackage{graphicx}
\usepackage{subfloat}
\usepackage{subfig}

\usepackage{diagbox}
\usepackage{bm}
\usepackage{amsfonts}
\usepackage{booktabs}
\usepackage{url}

\title{Robust Multi-agent Communication via  Multi-view Message Certification}
\author{%
  Lei Yuan\textsuperscript{\rm 1,2}, Tao Jiang\textsuperscript{\rm 1}, Lihe Li\textsuperscript{\rm 1}, Feng Chen\textsuperscript{\rm 1}, Zongzhang Zhang\textsuperscript{\rm 1}, Yang Yu\textsuperscript{\rm 1,2,}\thanks{Corresponding Author}\\
  \textsuperscript{\rm 1} National Key Laboratory for Novel Software Technology, Nanjing University, Nanjing, China \\
  \textsuperscript{\rm 2} Polixir.ai\\
  \texttt{\{yuanl, jiangt, lilh, chenf\}@lamda.nju.edu.cn, \{zzzhang, yuy\}@nju.edu.cn}
}
\date{}

\begin{document}

\maketitle

\begin{abstract} Many multi-agent scenarios require message sharing among agents to promote coordination, hastening the robustness of multi-agent communication when policies are deployed in a message perturbation environment. Major relevant works tackle this issue under specific assumptions, like a limited number of message channels would sustain perturbations, limiting the efficiency in complex scenarios. In this paper, we take a further step addressing this issue by learning a \textbf{ro}bust \textbf{M}ulti-\textbf{A}gent \textbf{C}ommunication 
policy via multi-view message \textbf{C}ertification, dubbed \textbf{CroMAC}. Agents trained under CroMAC can obtain guaranteed lower bounds on state-action values to identify and choose the optimal action under a worst-case deviation when the received messages are perturbed.
Concretely, we first model multi-agent communication as a multi-view problem, where every message stands for a view of the state. Then we extract a certificated joint message representation by a multi-view variational autoencoder (MVAE) that uses
a product-of-experts inference network. For the optimization phase, we do perturbations in the latent space of the state for a certificate guarantee. Then the learned joint message representation is used to approximate the certificated state representation during training. Extensive experiments in several cooperative multi-agent benchmarks validate the effectiveness of the proposed CroMAC.
\end{abstract}

\section{Introduction}
Many real-world problems are made up of multiple interactive agents, which could usually be modeled as a Multi-Agent Reinforcement Learning (MARL) problem~\cite{busoniu2008comprehensive, zhang2021multi}. Further, when the agents hold a shared goal, this problem refers to cooperative MARL~\cite{oroojlooyjadid2019review}, which shows great progress in diverse domains like power management~\cite{wang2021multi},  multi-UAV control~\cite{yun2022cooperative}, dynamic algorithm configuration~\cite{xue2022multi}, etc. Many methods are proposed to promote the coordination ability of MARL, including value-based methods~\cite{vdn,qmix,qplex}, policy-gradient-based methods~\cite{maddpg,mappo, ye2022towards}, and some variants~\cite{DBLP:conf/iclr/WangH0DZ21,cao2021linda,DBLP:conf/ijcai/YuanWWZCGZZY22,wen2022multiagent}, showing great progress in many complex and challenging benchmarks~\cite{papoudakis2021benchmarking,gorsane2022towards}. Nevertheless, prior works hugely depend on the strength of Deep Neural Networks (DNNs), whose vulnerability might cause catastrophic results when any perturbation happens~\cite{moos2022robust}. Recently this phenomenon has been tested in cooperative MARL~\cite{guo2022towards}, showing that a cooperative MARL system is of low robustness when encountering any perturbations (e.g., state, action, and reward).

Robustness has been widely investigated in single-agent reinforcement learning (RL)~\cite{moos2022robust}, and many works have applied different techniques to various aspects to investigate it. A prior popular way is to introduce an auxiliary adversary to play against the ego-system~\cite{pinto2017robust,zhang2020robust,vinitsky2020robust,song2022robust}, then model the process of policy learning as a minimax problem from the perspective of game theory~\cite{yu2021robust}, which may trigger performance deterioration or even unsafe behaviors when facing an unpredictable adversarial policy. Another kind of method tackles this issue by designing efficient and useful regularizers in the training process~\cite{zhang2020robust,oikarinen2021robust,sun2021strongest,wu2022robust}, showing efficient robustness in various domains. 
Certificate-based methods furthermore apply some techniques like vector-$\epsilon$-ball perturbations to obtain a certificate robustness guarantee during the training and testing phases~\cite{everett2021certifiable,wu2021crop,sun2022romax}. 
However, the MARL problem differs considerably from the single-agent setting, with multiple agents interacting with others~\cite{dorri2018multi}.

For the robustness of MARL, new challenges such as scalability~\cite{christianos2021scaling} arise as multiple agents interact with others in the training phase. For example, in the auxiliary adversary training paradigm, the action space of adversary policy may grow dramatically with respect to the number of agents in an MARL system. Works on robust MARL should then consider both the robustness and the multi-agent specificity. Some works design efficient mechanisms to obtain a robust policy to avoid overfitting to specific partners~\cite{van2020robust} or opponents~\cite{li2019robust}, and others consider the Markov decision process (MDP) itself to get a robust policy in response to state~\cite{zhou2022romfac}, reward~\cite{zhang2020robustmarl}, and action~\cite{hu2021robust, hu2022sparse}. Nonetheless, the robustness of a communication policy is much more complex~\cite{zhu2022survey}, as we should consider  \textbf{when} to give \textbf{what} perturbations on \textbf{which} message channel(s) to adversarially train the communication policy. Prior works mainly investigate the emergence of adversarial communication~\cite{blumenkamp2021emergence}, or impose constraints like a limited number of message channels~\cite{sun2022certifiably, xue2022mis} suffering from message perturbations. These approaches make progress somewhat, but the constraints hinder the robustness's completeness and are also far away from the real-world condition, as all the message channels could sustain perturbations~\cite{mackay2003information}. Even worse, these approaches lack formal robustness guarantees or certificates between each agent's received messages and decision-making.

With this in mind, we propose to promote robustness in current multi-agent communication methods. We posit obtaining a robust communication policy where \textbf{every messaging channel} could suffer from \textbf{perturbations} at \textbf{any time}. Specifically, for any agent in an $N$-agent system, it will receive $N-1$ messages. As each message is a different view of the state, we model the message-receiving process as a multi-view (a.k.a. ``multi-modal") problem, then obtain joint message representations with robustness guarantees from each received message by a Multi-view Variational AutoEncoder (MVAE) that uses a product-of-experts inference network. For the optimization phase, we first encode the state into a latent space, and do perturbations in this space to obtain a certificate relationship between the latent variable and the agents' $Q$-value.
Then, we train the message representation by approximating the certificated latent variables, and ensure certification between each message and the agents' $Q$-value implicitly. As we directly impose perturbations in the latent space, the problem of specific action designing for any auxiliary adversaries can be avoided. For evaluation, we conduct extensive experiments on various cooperative multi-agent benchmarks, including Hallway~\cite{ndq}, Level-Based Foraging~\cite{papoudakis2021benchmarking}, Traffic Junction~\cite{tarmac}, and two StarCraft Multi-Agent Challenge (SMAC) maps~\cite{ndq}. The results show that CroMAC achieves comparable or superior performance to multiple baselines. Moreover, visualization results show how CroMAC works, and more results demonstrate its high generality ability for different baselines on different conditions.

\section{Related Work} 
\paragraph{Multi-Agent Communication} plays a promising role in multi-agent coordination under partial observability, which considers \textbf{when} to communicate with \textbf{whom} and \textbf{what} contents to share~\cite{zhu2022survey}. The early relevant works mainly consider designing different communication paradigms to improve communication efficiency~\cite{foerster2016learning, communication16}. DIAL ~\cite{foerster2016learning} is a simple communication mechanism where agents broadcast messages to all teammates, allowing the gradient to flow among agents for end-to-end training with reinforcement learning. CommNet \cite{communication16} proposes an efficient centralized communication structure, where the outputs of the hidden layers from all the agents are collected and averaged to augment local observation. As the mentioned communication paradigm may cause message redundancy, some works employ techniques such as gate mechanisms~\cite{acml, i2c,ijcai2022p82} to explicitly decide whom to communicate with, or attention mechanisms~\cite{tarmac,doubleattention,wang2021tom2c} to weigh different messages. What messages to share among agents is another crucial issue. The most naive way is only to share local observations or their embeddings~\cite{foerster2016learning, ndq}, which inevitably causes bandwidth wasting or even degrades coordination efficiency. Towards a more efficient communication protocol, some methods utilize techniques like teammate modeling to generate more succinct and efficient messages~\cite{vbc,tmc,maic}. For the robustness of message sharing in CMARL, \citet{blumenkamp2021emergence} develop a new multi-agent learning model that integrates heterogeneous, potentially self-interested policies that share a differentiable communication channel to elicit the emergence of adversarial communications. \citet{xue2022mis} consider multi-agent adversarial communication, learning robust communication policy when some message senders are poisoned. A recent method named AME~\cite{sun2022certifiably} is proposed to acquire a robust communication policy when less than half of the agents in the system sustain noise and potential attackers.

\paragraph{Robustness in Single Agent Reinforcement Learning}~\citet{moos2022robust} involves perturbations that occur in different aspects in single agent reinforcement learning such as state, reward, policy, etc. Some prior methods introduce an adversary to achieve robustness via training the ego-system and the adversary in an alternative way~\cite{pinto2017robust,pan2019risk,vinitsky2020robust,zhang2020robust,song2022robust}. RARL~\cite{pinto2017robust} picks out specific robot joints that the adversary acts on to find an equilibrium of the minimax objective using an alternative learning adversary. RAP~\cite{vinitsky2020robust} and GC~\cite{song2022robust} improve RARL by learning population-based augmentation to the Robust RL formulation. However,  while these approaches provide better robust policies, it has been shown that such approaches can negatively impact policy performance in non-adversarial scenarios. Moreover, many unsafe behaviors may be exhibited during online attacks, potentially damaging the system controlled by the learning agent if adversarial training occurs in a physical rather than a simulated environment. Other methods improve robustness by designing useful and appropriate regularizers in the loss function~\cite{oikarinen2021robust,sun2021strongest,liangefficient}. \citet{zhang2020robust} formulate the problem of decision making under adversarial attacks on state observations as SA-MDP and learn a state-adversarial policy for multiple DRL methods like DDPG and DQN. RADIAL-RL~\cite{oikarinen2021robust} trains reinforcement learning agents with improved robustness against $l_p$-norm bounded adversarial attacks, showing superior performance on multiple benchmarks.The mentioned approaches achieve robustness compared to adversarial training, improving the sample efficiency as they need not train an auxiliary adversary. Furthermore, these mentioned methods lack theoretical guarantee, hastening some recent certificate robustness methods~\cite{qin2020learning,everett2021certifiable, wu2021crop,wu2021copa}. CARRL~\cite{qin2020learning} develops an online certifiably robust policy that computes guaranteed lower bounds on state-action values during execution to identify and choose a robust action under a worst-case deviation in input space due to possible adversaries or noise. CROP~\cite{wu2021crop} gives a solid theoretical guarantee for robust reinforcement learning and applies function smoothing techniques to train a robust policy. 

\paragraph{Multi-View (Modal) Representation Learning}
aims to learn feature representations from multi-view data using different views' information. Its main difficulty is to explicitly measure the content similarity between the heterogeneous samples. How to solve this problem roughly divides multi-view representation learning into three methods:
alignment representation~\cite{park2021learning}, joint representation~\cite{chen2012large}, as well as shared and specific representation~\cite{xu2020deep}. The key ideas of these methods are the same, which is establishing a common representation space by exploring the semantic relationship among the multi-view data. One popular and promising way is to use generative models like VAE~\cite{DBLP:journals/corr/KingmaW13}, which generate this representation space in two ways: cross-view generation and joint-view generation. The former learns a conditional generative model over all views by applying techniques like conditional VAE~\cite{sohn2015learning}. Nevertheless, the latter learns the joint distribution of the multi-view data. For example,  MVAE~\cite{wu2018multimodal} models the joint posterior as a product-of-experts (POE), and JMVAE~\cite{suzuki2016joint} learns a shared representation with a joint encoder. Please refer to~\cite{yan2021deep, bayoudh2022survey} for a comprehensive review. After the representation space is established by multi-view learning, some approaches use it to solve the modality missing problem~\cite{ma2021smil}, or obtain a compact representation from incomplete views~\cite{xu2015multi}. \citet{li2019multi} extend the partially observable Markov decision processes (POMDPs) to support more than one observation model and propose two solutions through observation augmentation and cross-view policy transfer in a reinforcement learning problem. DRIBO~\cite{fan2022dribo} leverages the sequential nature of RL to learn robust representations that encode only task-relevant information from observations based on the unsupervised multi-view setting. \citet{kinose2022multi} introduce a novel reinforcement learning agent for integrated recognition and control from multi-view observations. To the best of our knowledge, none of any MARL approaches use multi-view learning to train the communication policy. We take a further step in this direction to get a robust message representation.

\paragraph{Multi-agent Robustness}
Robustness also plays a promising role in MARL~\cite{guo2022towards}, but suffer from extra challenges that do not appear in the single-agent setting, as interactions exist among agents~\cite{gronauer2022multi}, leading to new and specific considerations such as non-stationa\-rity~\cite{papoudakis2019dealing}, credit assignment~\cite{wang2021towards}, and scalability~\cite{christianos2021scaling} when improving the robustness of any multi-agent system~\cite{guo2022towards}. One type of relevant work aims to investigate the robustness of a learned coordination policy. \citet{lin2020robustness} first learn an observation attacker via RL, then use it to poison one manually selected agent, showing the multi-agent system is vulnerable to 
observation perturbation. \citet{guo2022towards} recently do more comprehensive robustness testing on reward, state, and action for typical MARL methods like QMIX~\cite{qmix} and MAPPO~\cite{mappo}. As for robustness improvement in MARL, research is conducted on multiple aspects. Many prior works focus on designing an efficient approach to learning a robust coordination policy to avoid overfitting to specific partners~\cite{van2020robust} or opponents~\cite{li2019robust}. Akin to considering the MDP in a single-agent setting (e.g., state, reward, action), R-MADDPG~\cite{zhang2020robustmarl} considers the model uncertainty of an MARL system, then introduces the concept of robust Nash equilibrium. \citet{hu2021robust} apply a heuristic rule to investigate the robustness of MARL when some agents suffer from action mistakes, and utilize correlated equilibrium theory to learn a robust coordination policy. Robustness in multi-agent communication has also attracted some attention in recent years. \citet{mitchell2020gaussian} apply a filter based on the Gaussian process to extract valuable content from noisy messages. \citet{tu2021adversarial} study robustness at the neural network level for secure multi-agent systems. \citet{xue2022mis} model multi-agent communication as a two-player zero-sum game and apply the policy-search response-oracle (PSRO) technique to learn a robust communication policy. The most related work to ours is Ablated Message Ensemble (AME)~\cite{sun2022certifiably}, which assumes no more than half of the message channels in the system may be attacked, then introduces an ensemble-based defense method to achieve robustness. However, we will show that this approach performs poorly in complex scenarios, as the constraints may impede robust efficiency.


\section{Problem Formulation}\label{Sec.3}

We consider a fully cooperative MARL communication problem, which can be formally modeled as a Decentralized Partially Observable MDP under Communication (Dec-POMDP-Com)~\cite{xue2022mis} and formulated as a tuple $\langle \mathcal{N}, \mathcal{S}, \mathcal{A}, P, \Omega, O, R, \gamma , \mathcal{M} \rangle$, where $\mathcal{N} = \{1, \dots, n\}$, $\mathcal{S}$, $\mathcal{A}$, and $\Omega$ are the sets of  agents, states, actions, and observations, respectively. $O$ is the observation function,  $P$ denotes the transition function, 
$R$ represents the reward function, $\gamma \in [0, 1)$ stands for the discounted factor, and $\mathcal{M}$ indicates the message set. Due to partially observable nature of the environment,  each agent $i \in \mathcal{N}$ can only 
obtain the local observation $o_i\in \Omega$, and hold an individual policy $\pi(a_i\mid \tau_i,m_i)$, where $\tau_i$ represents the output of a trajectory encoder (e.g., GRU~\cite{gru}) which encodes $(o_i^1, a_i^1, \dots, o_i^{t-1}, a_i^{t-1}, o_i^t)$, and $m_i \in \mathcal{M}$ is the messages received by agent $i$ and $m_{ij}$ represents the message transmitted from $j$ to $i$. As each agent can behave as a message sender as well as a message receiver, this paper considers learning useful message representation on the receiving end, and agents only use local information (e.g., $\tau_i$, and we use $m_{:,i}$ for generality) as message $m_{:,i}$ to share within the team. We aim to find an optimal policy under the setting where each message channel in the multi-agent system may suffer from perturbations. In line with the widely used state-adversarial MDP (SA-MDP) in single-agent RL~\cite{zhang2020robust,qiaoben2021understanding}, we formulate this setting as a message-adversarial Dec-POMDP-Com (MA-Dec-POMDP-Com).

In an MA-Dec-POMDP-Com, we introduce a message adversary $v(m):m \rightarrow{\hat{m}}$. The 
adversary perturbs the messages received by each agent, such that agent $i$ takes action by $\pi(a_i|\tau_i,\hat{m}_i)$. The joint action $\boldsymbol{a}=\langle a_1, \dots, a_n \rangle$ leads to the next state $s'\sim P( \cdot \mid s, \boldsymbol{a})$ and the global reward $R(s, \boldsymbol{a})$. The formal objective is to find a joint policy $\boldsymbol{\pi}(\boldsymbol{\tau}, \boldsymbol{a})$ to maximize the global value function $Q_{\rm tot}^{\boldsymbol{\pi}}(\boldsymbol{\tau}, \boldsymbol{a}) =\mathbb{E}_{s,\boldsymbol{a} }\left[\sum_{t=0}^\infty\gamma^tR(s, \boldsymbol{a})\mid s_0=s, \boldsymbol{a_0}=\boldsymbol{a}, \boldsymbol{\pi}\right]$, with $\boldsymbol{\tau}=\langle \tau_1, \dots, \tau_n \rangle$. 
If the adversary can perturb a message $m$ arbitrarily without bounds, the problem becomes trivial. To
fit our method to the most realistic settings, we restrict the power of an adversary to a
perturbation set $\mathcal{B}$, i.e., $\mathcal{B}=\{\hat{m} \mid \|m-\hat{m}\|_p \le \epsilon \}$, where $\epsilon$ is the given perturbation magnitude and $p$ determines the type of norm. Our experiments in this paper focus on $p=\infty$.
Furthermore,  since $\mathcal{B}(m)$ is usually a small set nearby $m$, 
our adversary applies FGSM~\cite{goodfellow2014explaining} to learn a perturbation vector $\Delta$, and we project $m+\Delta$ to $\mathcal{B}(m)$.

\begin{figure*}
  \centering
  \includegraphics[width=1\textwidth]{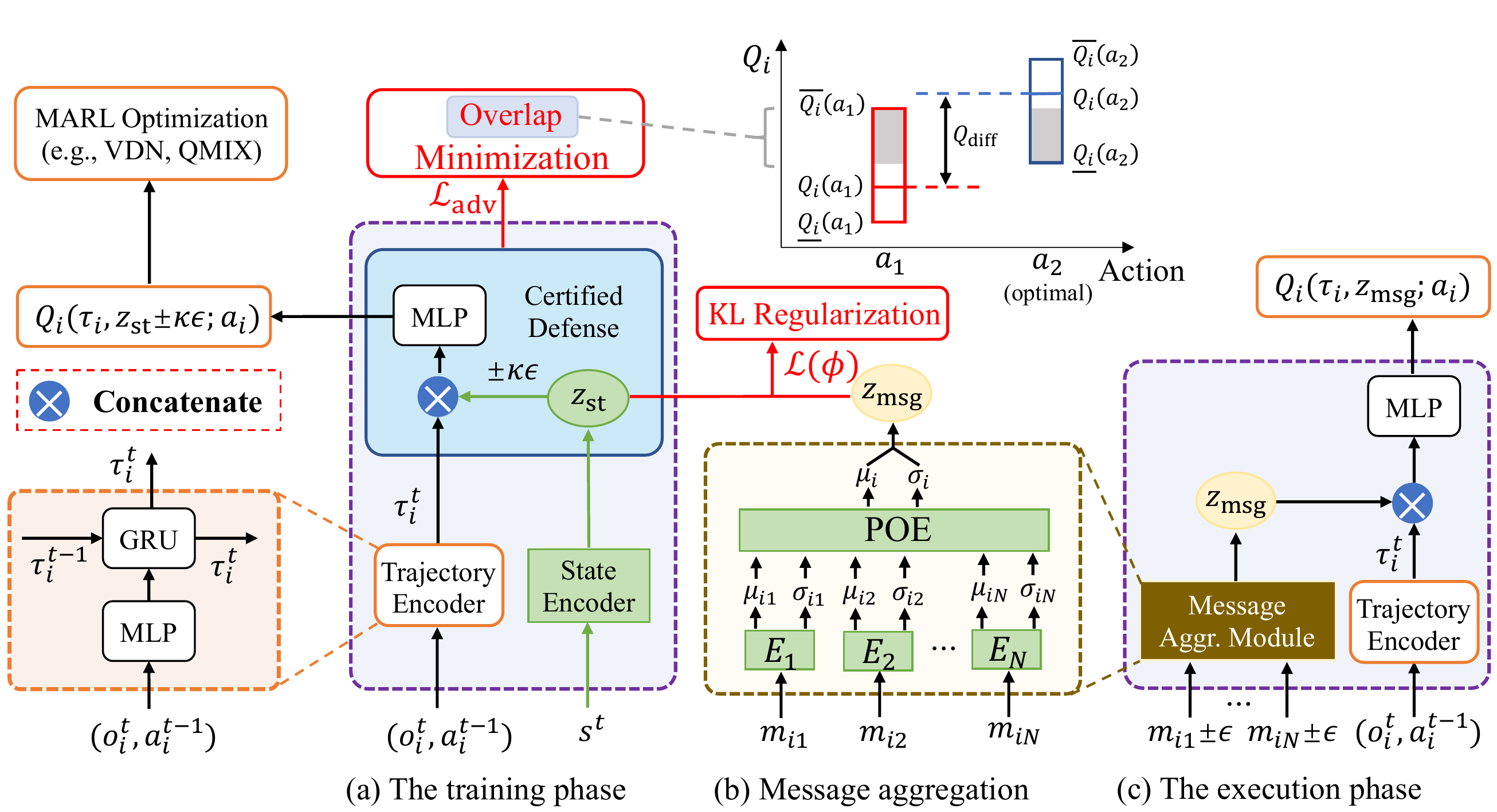}
  \caption{Structure of CroMAC. (a) During the training phase, we encode the state into latent variables $z_{\rm st}$, then perturb it to gain a certificate guarantee between $z_{\rm st}$ and  $Q_i(\tau_i,z_{\rm st}\pm \kappa \mathbb{\epsilon}; a_i)$, and this process is optimized via
  minimizing the overlap between the output bounds of action values to get a large difference in the outcome. The whole process can be optimized by any value decomposition methods like QMIX~\cite{qmix}, and the output of the message aggregation module $z_{\rm msg}$ is then used to approximate $z_{\rm st}$ by minimizing their distance (e.g., KL divergence). (b) The message aggregation module. Each message $m_{ij}$ is encoded into a latent space via a message encoder $E_j$, where $j\in \{1, \cdots, i - 1, i + 1, \cdots, N\}$, and the parameters of $E_j$ are regularized to obtain certificates between the joint message representation and each message. (c) After training, we use the learned message aggregation module
  and other shared modules like trajectory encoder
  to make a decision in a decentralized way.} 
  \label{cCroMACframework}
\end{figure*}

\section{Method}
This section gives a detailed description of our proposed CroMAC (Fig.~\ref{cCroMACframework}), a novel approach that achieves a robust communication policy under MA-Dec-POMDP-Com. As each message is a view of the state, we first apply a multi-view variational autoencoder (MVAE) that uses a product-of-experts inference network to extract a joint message representation.
Then we obtain the bounds between the joint message representation and each message by interval bound propagation~\cite{ibp}.
In the training phase, we first encode state $s_t \in \mathcal{S}$ into a latent variable $z_{\rm st}$, then we impose perturbations in the latent space to gain a certificate guarantee between $z_{\rm st}$ and each state-action value $Q_i(\tau_i,z_{\rm st}\pm \kappa \mathbb{\epsilon}; a_i)$, where $\pm \kappa\mathbb{\epsilon}$ represents that the variable suffers from $\ell_{\infty}$-norm perturbations within budget $\kappa\mathbb{\epsilon}$ and $\kappa$ is a constant. Finally, the joint message representation $z_{\rm msg}$ is optimized by approximating $z_{\rm st}$ via minimizing the  Kullback-Leibler divergence between these two variables, endowing certification between each message and each state-action value implicitly. In the execution phase, we only use the message aggregation module and the trajectory encoder to make decisions in a decentralized way.  


\subsection{Multi-view Multi-agent Communication}
We consider learning a robust communication policy in  an MA-Dec-POMDP-Com. For each agent $i$, there are $N-1$ message channels,
thus each agent receives multiple available messages about the environment. Inspired by the widely used multi-view learning~\cite{li2018survey,hwang2021multi}, we apply the product-of-experts (POE)~\cite{gowal2018effectiveness} technique to extract joint message representations.
Formally,  agent $i$ receives multiple messages $m_{ij}^t$ from teammate $j\in \{1, \cdots, i - 1, i + 1, \cdots, N\}$ and holds its local history $\tau_i^t$ as $m_{ii}^t$ at time $t$. We assume each message is conditioned on an unknown hidden variable $z_{ij}^t$, then the generation of multiple messages can be modeled as a multi-view variational autoencoder process. We then optimize the Evidence Lower BOund (ELBO) to maximize the marginal likelihood with a message encoder
$q_{\bm \phi_{\rm enc}}(z_{ij}^t|m_{ij}^t)$ parameterized with $\bm \phi_{\rm enc}$, and a message decoder $p_{\bm \phi_{\rm dec}}(m_{ij}^t|z_{ij}^t)$ with parameter $\bm \phi_{\rm dec}$:
\begin{align}
	\operatorname{ELBO}(m_{ij}^t) 
	\triangleq  
	\mathbb{E}_{q_{
	\bm \phi_{\rm enc}}(z_{ij}^t \mid m_{ij}^t)}\left[  \log p_{\bm \phi_{\rm dec}}\left(m_{ij}^t\mid z_{ij}^t\right)\right]
     - \operatorname{KL}\left[q_{\bm \phi_{\rm enc}}(z_{ij}^t \mid m_{ij}^t), p(z_{ij}^t)\right],
     \label{mvae}
\end{align}
where $\text{KL}[p,q]$ is the Kullback-Leibler 
divergence between distributions $p$ and $q$. The first term in Eqn.~\ref{mvae} is the reconstruction likelihood, and the second term aims to guarantee that the output of the encoder is similar to the prior distribution
$p(z_{ij}^t)$, and can be regarded as a regularization term. The message encoder $q_{\bm \phi_{\rm enc}}$ outputs parameters of an $n$-multivariate Gaussian distribution $\mathcal{N}({ \mu}^t_{ij},  {\sigma}^t_{ij})$, where ${\mu}^t_{ij}$ and ${\sigma}^t_{ij}$ are the mean and standard deviation of  $p(z_{ij}^t)$, respectively.
As all messages $\{m_{i1}^t,\cdots,m_{iN}^t\}$ are conditionally independent given the common latent variable $z_i^t$, we assume a generative model for all the messages in the
form: 
\begin{align}
&p(m_{i1}^t,\cdots,m_{iN}^t,z_i^t)= p(z_i^t)p(m_{i1}^t|z_i^t) p(m_{i2}^t|z_i^t)\cdots p(m_{iN}^t|z_i^t). 
\end{align}
Then Eqn.~\ref{mvae} can be extended as:
\begin{align}
	     \operatorname{ELBO}(m_i^t) \triangleq  \mathbb{E}_{q_{\bm \phi_{\rm enc}}(z_i^t \mid m_i^t)}\left[\sum_{j=1}^N  \log p_{\bm \phi_{\rm dec}}\left(m_{ij}^t\mid z_i^t\right)\right]
     - \operatorname{KL}\left[q_{\bm \phi_{\rm enc}}(z_i^t \mid m_i^t), p(z_i^t)\right],
     \label{mvaevari}
\end{align}

where $m_i^t=\{m_{i1}^t,\cdots,m_{iN}^t\}$ is the set of messages agent $i$ receives at time $t$ and its local history $m_{ii}^t$ (i.e., $\tau_i^t$). Then, this message generation process can be treated as a multi-view representation learning problem~\cite{suzuki2016joint}. 
We use the inference network $q(z_i^t \mid m_i^t)$ as a variational distribution to approximate the true posterior $p(z_i^t \mid m_i^t)$, then get the relationship among the joint- and single- view posteriors as: 
\begin{equation}
\begin{aligned}
    p(z_i^t\mid m_i^t) &= \frac{p(m_i^t\mid z_i^t)p(z_i^t)}{p(m_i^t)} = \frac{p(z_i^t)}{p(m_i^t)}\prod_{j=1}^N p(m_{ij}^t\mid z_i^t) \\
    &= \frac{p(z_i^t)}{p(m_i^t)}\prod_{j=1}^N \frac{p(z_i^t\mid m_{ij}^t)p(m_{ij}^t)}{p(z_i^t)}  \\
    &= \frac{\prod_{j=1}^N p(m_{ij}^t)}{p(m_i^t)} \frac{\prod_{j=1}^N p(z_i^t\mid m_{ij}^t)}{\prod_{j=1}^{N-1} p(z_i^t)}   \\
    & \propto \frac{\prod_{j=1}^N p(z_i^t\mid m_{ij}^t)}{\prod_{j=1}^{N-1} p(z_i^t)}  
     \approx \frac{\prod_{j=1}^N [{q}(z_i^t\mid m_{ij}^t)p(z_i^t)]}{\prod_{j=1}^{N-1} p(z_i^t)} \\
    &= p(z_i^t)\prod_{j=1}^N {q}(z_i^t\mid m_{ij}^t). 
    \label{eqn.poe2}
\end{aligned}
\end{equation}
The last two lines in Eqn.~\ref{eqn.poe2} hold as we use $q(z_i^t\mid m_{ij}^t)p(z_i^t)$ to approximate $p(z_i^t\mid m_{ij}^t)$ so that the inference network is composed of $N$ neural networks ${q}(z_i^t\mid m_{ij}^t)$, and if each view is homogeneous, we can even replace them with only one network with shared parameters. Akin to the standard VAE~\cite{DBLP:journals/corr/KingmaW13}, we apply a deep neural network (e.g., MLP) to model the message encoder $q_{\bm \phi_{\rm enc}}(z_i^t|m_{ij}^t)$, which outputs the parameters of the Gaussian distribution $\mu_{ij},\sigma_{ij}^2$.
As for now, we can combine the multiple outputs of message encoder in a simple analytical way: a product
of Gaussian experts is itself Gaussian~\cite{cao2014generalized}:
\begin{equation}
    \begin{aligned}
    \mu_i &= (\sum_{j=1}^N \mu_{ij}\mathrm{T}_{ij})(\sum_{j=1}^N \mathrm{T}_{ij})^{-1}, \\
    \sigma_i^2 &= (\sum_{j=1}^N \mathrm{T}_{ij})^{-1},
    \label{mvaeprof}
\end{aligned}
\end{equation}
where $\mu_i$
and $\sigma_{i}^2$ are the mean and variance of the learned joint message representation's Gaussian distribution, and $\mu_{ij}$ and $\sigma_{ij}^2$ are the mean and variance of the $i$-th agent's $j$-th Gaussian distribution through message encoder, $\mathrm{T}_{ij}=(\sigma_{ij}^2)^{-1}$ is the inverse of the variance. The detailed derivative process can be seen in~\ref{poeprof}. 

\subsection{Message  Certificates via  Bound Propagation} \label{Sec4.2}
Though we have combined all the received messages into a joint message representation, the learned joint message representation still lacks a certificated guarantee with each received message under perturbation. In this part, 
we aim to achieve this using the interval bound propagation technique. 
Formally, consider agent $i$ receives messages $m_i=\{m_{i1},\cdots,m_{iN}\}$ under perturbations of $\ell_{\infty}$-norm attack within given budget $\epsilon$, the upper and lower bounds are $\overline{m_i}=\{\overline{m_{i1}}, \cdots, \overline{m_{iN}}\}=\{m_{i1}+\boldsymbol{\epsilon}, \cdots, m_{iN}+\boldsymbol{\epsilon\}}$ and $\underline{m_i}=\{\underline{m_{i1}}, \cdots, \underline{m_{iN}}\}=\{m_{i1}-\boldsymbol{\epsilon}, \cdots, m_{iN}-\boldsymbol{\epsilon\}}$, respectively. The 
averages and residuals of the upper and lower bounds can be denoted as:
\begin{equation}
    \begin{aligned}
    \hat{\mu}_0&=\frac{1}{2}(\overline{m_i}+\underline{m_i})=m_i, \\
    \hat{r}_0&=\frac{1}{2}(\overline{m_i}-\underline{m_i})=\boldsymbol{\epsilon}.
\end{aligned}
\end{equation}
%
Here, we use ``average" instead of ``mean" to distinguish it from the one in VAE, and $\hat{\mu}$ and $\hat{r}$ are used to represent averages and residuals, respectively. 
For simplification of notation and mathematical derivation, 
 we assume each message encoder has only one layer fully-connected network with shared parameters, and we can use any NNs with arbitrary depths and types (linear
or non-linear) by getting a reasonable bound propagation mechanism~\cite{gowal2018effectiveness}.
 Further, we here use $W_m,b_m$, and $W_v,b_v$ to represent the parameters of the two separate fully connected layers in the message encoder, which output the means and variances of the Gaussian distributions, respectively. Thus we can propagate the bounds through the one MLP layer by matrix multiplication. The averages and residuals of the bounds come to be:
\begin{equation}
    \begin{aligned}
    \hat{\mu}_m&=\{W_m\hat{\mu}_0(1)+b_m,\cdots,W_m\hat{\mu}_0(N)+b_m\},\\
    \hat{r}_m&=\{|W_m|\hat{r}_0(1),\cdots,|W_m|\hat{r}_0(N)\},\\
    \hat{\mu}_v&=\{W_v\hat{\mu}_0(1)+b_v,\cdots,W_v\hat{\mu}_0(N)+b_v\},\\
    \hat{r}_v&=\{|W_v|\hat{r}_0(1),\cdots,|W_v|\hat{r}_0(N)\},
\end{aligned}
\end{equation}
where $(\hat{\mu}_m, \hat{r}_m)$ and $(\hat{\mu}_v, \hat{r}_v)$ stand for the $($average, residual$)$ pairs of the mean outputs' bounds and the variance outputs' bounds, respectively, and $| \cdot |$ is the element-wise absolute value operator. We use ReLU as the activation function which is element-wise monotonic so that we can omit it as it will not affect the propagating bounds. Thus the upper and lower bounds of each single message representation (i.e., mean and variance) can be written as:
%
\begin{equation}
\begin{aligned}
    \overline{z_m}&=\hat{\mu}_m+\hat{r}_m=\{W_mm_{i1}+b_m+|W_m|\boldsymbol{\epsilon},\cdots,W_mm_{iN}+b_m+|W_m|\boldsymbol{\epsilon}\},
    \\
    \underline{z_m}&=\hat{\mu}_m-\hat{r}_m=\{W_mm_{i1}+b_m-|W_m|\boldsymbol{\epsilon},\cdots,W_mm_{iN}+b_m-|W_m|\boldsymbol{\epsilon}\},
    \\
    \overline{z_v}&=\hat{\mu}_v+\hat{r}_v=\{W_vm_{i1}+b_v+|W_v|\boldsymbol{\epsilon},\cdots,W_vm_{iN}+b_v+|W_v|\boldsymbol{\epsilon}\}, \\
    \underline{z_v}&=\hat{\mu}_v-\hat{r}_v=\{W_vm_{i1}+b_v-|W_v|\boldsymbol{\epsilon},\cdots,W_vm_{iN}+b_v-|W_v|\boldsymbol{\epsilon}\}.
\end{aligned}
\end{equation}

 Consider the relationship in Eqn.~\ref{mvaeprof}, we then have:
\begin{equation}
\begin{aligned}
    &Z_M = \big(\sum_i z_m(i)z_v(i)^{-1}\big)\big(\sum_i z_v(i)^{-1}\big)^{-1}, \\
    &Z_V = \big(\sum_i z_v(i)^{-1}\big)^{-1},
\end{aligned}
\label{eqn.dsfds}
\end{equation}
where $Z_M, Z_V$ represent the mean and variance of the joint message representation. However,  we cannot get the upper and lower bounds as the POE we use is not affine neural layers. Notice that $Z_V$ is actually the Harmonic Mean~\cite{gautschi1974harmonic} of $\frac{z_v(i)}{N}$ (the Harmonic Mean of variables $x_i$ is $H_n=\frac{n}{\frac{1}{x_1}+\frac{1}{x_2}+\cdots+\frac{1}{x_n}}$ ) while $Z_M$ is the Weighted Harmonic Mean of $z_m(i)$ with weights $\frac{z_m(i)}{z_v(i)}$ (the Weighted Harmonic Mean of $x_i$ with weights $w_i$ is $H_n=\frac{m_1+m_2+\cdots+m_n}{\frac{m_1}{x_1}+\frac{m_2}{x_2}+\cdots+\frac{m_n}{x_n}}$). Here, variances $z_v(i)$ are non-negative and means $z_m(i)$ can be normalized to a positive range, 
therefore we can scale Eqn.~\ref{eqn.dsfds} appropriately by the properties of Harmonic Mean to infer the upper and lower bounds.
We here prove one of them for simplicity. Take the variance term for example and simplify  $W_v,b_v,Z_V,z_v$ as $W,b,Z,z$, we have:
\begin{equation}
\begin{aligned}
    \overline{Z} &= \big(\sum_i \overline{z}(i)^{-1}\big)^{-1}\le \frac{\mathbf{1}\cdot \max(\overline{z})}{N}, \\
    \underline{Z} &= \big(\sum_i \underline{z}(i)^{-1}\big)^{-1} \ge \frac{\mathbf{1}\cdot \min(\underline{z})}{N}.
    \label{eqn.vl} 
\end{aligned}
\end{equation}
Note that $\mathbf{1}$ is an all-1-vector with the same dimension as $\overline{Z}$. $(\le,\ge)$ signs here act
on each element of vectors, and $(\max,\min)$ operations find the maximum or minimum number of vectors of the set. We can get the upper bound of the final integration error:
\begin{equation}
\begin{aligned}
   \max(\overline{Z}-{Z}_{\rm true},{Z}_{\rm true}-\underline{Z})&\le \overline{Z} -\underline{Z}  
    \le \frac{\mathbf{1}\cdot (\max(\overline{z})-\min(\underline{z})) }{N},\label{eqn.vi} 
\end{aligned}
\end{equation}
where ${Z}_{\rm true}$ stands for the ground truth value. Assume the $p$-th element of the $j$-th vector of $\overline{z}$ is $\max(\overline{z})$ and $q$-th element of the $k$-th vector of $\underline{z}$ is $\min(\underline{z})$, we get
\begin{equation}
\begin{aligned}
   \max(\overline{z})-\min(\underline{z})
   &=(Wm_{ij}+b+|W|\boldsymbol{\epsilon})_p-(Wm_{ik}+b-|W|\boldsymbol{\epsilon})_q  \\
   &=W_{p,:}m_{ij}-W_{q,:}m_{ik}+b_p-b_q+(|W|_{p,:}-|W|_{q,:})\boldsymbol{\epsilon}, 
\end{aligned}
\end{equation}
where $W_{p,:}$ means the $p$-th row of matrix $W$. We can notice that the integration error can be limited to a constant $\Vert |W|_{p,:}-|W|_{q,:}\Vert_1/N$ times $\boldsymbol{\epsilon}$ if $W,b,m_i$ are bounded. Through subsequent experiments, we found that good robustness could be achieved when $\kappa$ times the noise is considered in the integrated information with only $W$ bounded to $[C\_MIN,C\_MAX]$, here $C\_MIN,C\_MAX,\kappa$ are hyperparameters and we let $C\_MIN=-C\_MAX$. 
\subsection{Robustness Training Scheme}
As we have obtained the theoretical guarantee between the received messages and the learned joint message representation, now this subsection describes how to acquire a robust communication policy. Following the popular Centralized Training and Decentralized Execution (CTDE) paradigm~\cite{ctde2016,lyu2021contrasting}, during the training phase,  we  use a state encoder (e.g., additional VAE~\cite{DBLP:journals/corr/KingmaW13}) to encode the state $s$ into a latent space with parameter $\bm \psi$:
\begin{equation}
\begin{aligned}
\mathcal{L}(\bm \psi)=
-\mathbb{E}_{q_{\bm \psi_{\rm enc}}({\bm z_{\rm st}}\mid s)}[\log {p_{\bm \psi_{\rm dec}}(s \mid  z_{\rm st})}] 
+ \operatorname{KL}\left[q_{\bm \psi_{\rm enc}}({\bm z}_{\rm st}\mid s), p(\bm z_{\rm st})\right],
\end{aligned}
\label{loss1}
\end{equation}
where the operators are similar to Eqn.~\ref{mvae}. We can then apply any robust single-agent RL algorithm to achieve robustness. We here implement our method on RADIAL-RL~\cite{oikarinen2021robust} as it principles a framework in adversarial training with strong theoretical guarantee and robustness performance. 
Then we can minimize the following loss to optimize each agent's individual policy:
\begin{align}
    \mathcal{L}_{\rm adv}= \mathbb{E}_{(s,\boldsymbol{a},s',r)}\left[\sum_i\sum_y Q_{\rm diff}^i(\tau,z_{\rm st};y)\cdot Ovl^i(\tau,z_{\rm st},\kappa\epsilon; y)  \right], \label{eqn.advloss}
\end{align}
with \begin{align}
    &Q_{\rm diff}^i(\tau,z_{\rm st};y)=\max(0,Q^i(\tau,z_{\rm st};a)-Q^i(\tau,z_{\rm st};y)), \nonumber \\
    &Ovl^i(\tau,z_{\rm st},\kappa\epsilon;y)=\max(0,\overline{Q}^i(\tau,z_{\rm st},\kappa\epsilon;y)-\underline{Q}^i(\tau,z_{\rm st},\kappa\epsilon;a)) \nonumber,\\
\end{align}
where $i$ is the identification of each agent, $y$ is each action, $a$ is the chosen action, and $\overline{Q},\underline{Q}$ can be computed by interval bound propagation under $\ell_{\infty}$-norm perturbation within budget $\kappa \mathbb{\epsilon}$, which is readily available as there is only MLP network existing between the $Q$-values and $(\tau,z_{\rm st})$.  $Ovl$ represents the overlap between the bounds of two actions which can be seen in Fig.~\ref{cCroMACframework}, and $Q_{\rm diff}$ measures the relative quality between two actions as we can ignore the overlap if they are similar enough. We note that the model's initial training will be hindered if we add the robust loss. Therefore, it is better to start robust training after the training is stable, and we use $T_r$ to control it. Then we optimize the joint message representation by minimizing the $\operatorname{KL}$ 
divergence between $z_{\rm st}$ and $z_{\rm msg}$, and we use only the message encoder $\bm \phi_{\rm enc}$ to make the inference of the joint message representation:
\begin{align}
    \mathcal{L}(\bm \phi) = \operatorname{KL}\left[sg(z_{\rm st}),q_{\bm \phi_{\rm enc}}(z_{\rm msg} \mid m)\right], \label{loss.2}
\end{align}
 where $sg(\cdot)$ denotes gradient stop, and $z_{\rm msg}$ is the joint message representations sampled from $\mathcal{N}({ \mu}_{i},  {\sigma}_{i}^2)$. Then the overall loss function is as follows:
\begin{equation}
\begin{aligned}  \label{loss_tot}
    \mathcal{L} = \mathcal{L}_{\rm TD}(\bm\theta) + \alpha_{1}\mathcal{L}(\bm \psi)+ \alpha_{2}\mathcal{L}(\bm \phi) +\mathbb{I}(t>T_r) \alpha_{3}\mathcal{L}_{\rm adv}.
\end{aligned}
\end{equation}
Here, $\mathcal{L}_{\rm TD}( \bm\theta)$ is the temporal difference loss, $\alpha_{1}$, $\alpha_{2}$, and $\alpha_{3}$ are adjustable hyperparameters for each loss function accordingly, and $\mathbb{I}(\cdot)$ is the indicator function. In the CTDE framework, the mixing network will be removed during the decentralized execution phase. 
To prevent the lazy-agent problem~\cite{vdn} and reduce model complexity, we make the local network have the same parameters for all agents. The pseudo-code is shown in~\ref{algorithm}.

\begin{figure}[htbp]
    \centering
    \includegraphics[width=\linewidth]{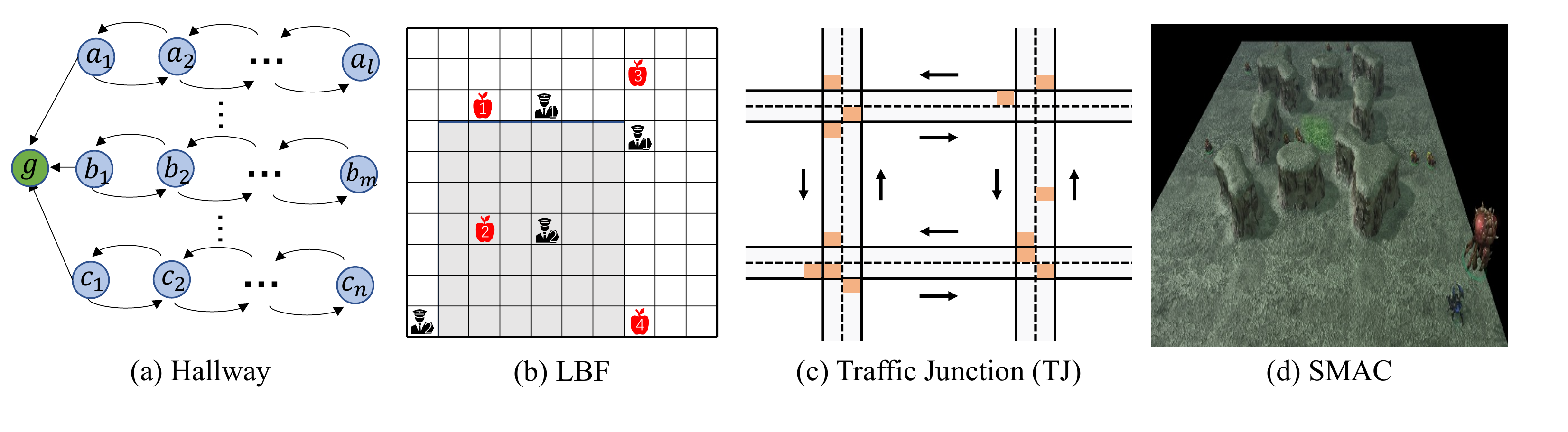}
	\caption{Multiple benchmarks used in our experiments.}
	\label{fig:envapp}
\end{figure}
\section{Experimental Results}
In this section, we design experiments on multiple complex scenarios to evaluate the communication robustness for the following questions: (1) How is the robustness of our proposed method on multiple benchmarks compared with baselines (Sec.~\ref{resultandanalysis}) ? (2) What is the generalization ability of CroMAC when encountering different message perturbations  (Sec.~\ref{generalization}) ? (3) Can CroMAC be integrated into multiple cooperative MARL methods in different communication conditions, and how does each hyperparameter influence its performance (Sec.~\ref{sensitive}) ?

For empirical evaluation, we compare CroMAC with multiple baselines on different cooperative tasks, including Hallway~\cite{ndq}, Level-Based Foraging (LBF)~\cite{papoudakis2021benchmarking}, Traffic Junction (TJ)~\cite{tarmac}, and two maps from StarCraft Multi-Agent Challenge (SMAC)~\cite{ndq}. CroMAC is implemented on QMIX if not specified based on PyMARL\footnote{ {https://github.com/oxwhirl/pymarl}}. All results are illustrated with mean performance
and standard error on 5 random seeds. Detailed
network architecture and hyperparameter choices are
shown in~\ref{detailedimplement}.

\subsection{Baselines and Environments}
We consider multiple baselines with different communication ability, where QMIX~\cite{qmix} is a value-based baseline, where no message sharing among agents,  showing excellent performance on diverse
multi-agent benchmarks~\cite{papoudakis2021benchmarking}.
AME~\cite{sun2022certifiably} is a recent strong method for the robustness of multi-agent communication, which assumes no more than half of the agents may suffer from message perturbations, and an ensemble-based defense approach is then introduced to realize the robustness goal. Full-Comm adopts a full communication paradigm, where each agent receives messages from all teammates at each timestep without message perturbations both in the training and testing phases, which can be seen as an upper-bound performance algorithm. For the ablation studies, we consider multiple variants of CroMAC.  CroMAC w/o robust and CroMAC w/o adv are two variants of our proposed CroMAC, the former does not have the proposed robust training scheme while the latter is conducted
in the non-perturbed condition in both the training and testing phases.
We consider multiple benchmarks,  where Hallway~\cite{ndq} is a cooperative environment under partial observability,  with $m$ agents are randomly initialized at different positions and required to arrive at the goal $g$ simultaneously. We consider two scenarios with 
different numbers of agents and lengths of the hallway. LBF~\cite{papoudakis2021benchmarking} is another cooperative, partially observable grid world game where agents should coordinate to collect food concurrently. As the original version focuses on exploration, here we modify it by making only one agent able to observe the map, which needs strong communication to complete this task. TJ~\cite{tarmac} is another popular benchmark used to test communication ability, where multiple cars move along two-way roads with one or more road junctions following predefined routes. We test on the modified slow and fast scenarios where different maps have a different probability of adding new cars. Two maps named {1o\_2r\_vs\_4r} and {1o\_10b\_vs\_1r} from SMAC~\cite{ndq} that require efficient communication are also used to test the robustness in more complex scenarios. Details are presented in~\ref{benchmarkdt}.

\subsection{Robustness Comparison and Analysis} \label{resultandanalysis}
\begin{figure*}
	\centering
\subfloat[Hallway: 4x5x6]{
    \label{hallway1}
\includegraphics[width=0.24\textwidth]{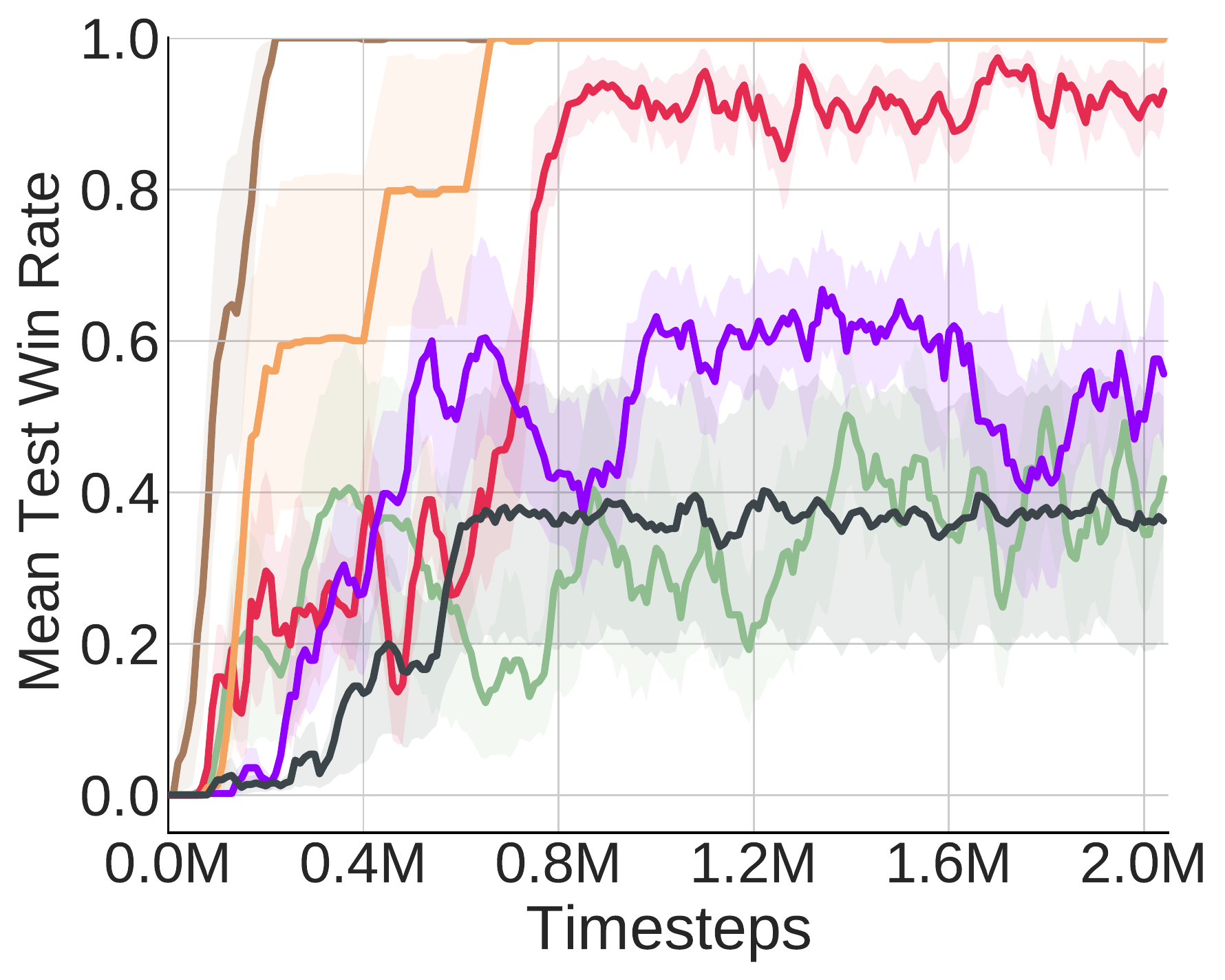}
    }
\subfloat[LBF: 3p-1f]{
    \label{lbf1}
\includegraphics[width=0.225\textwidth]{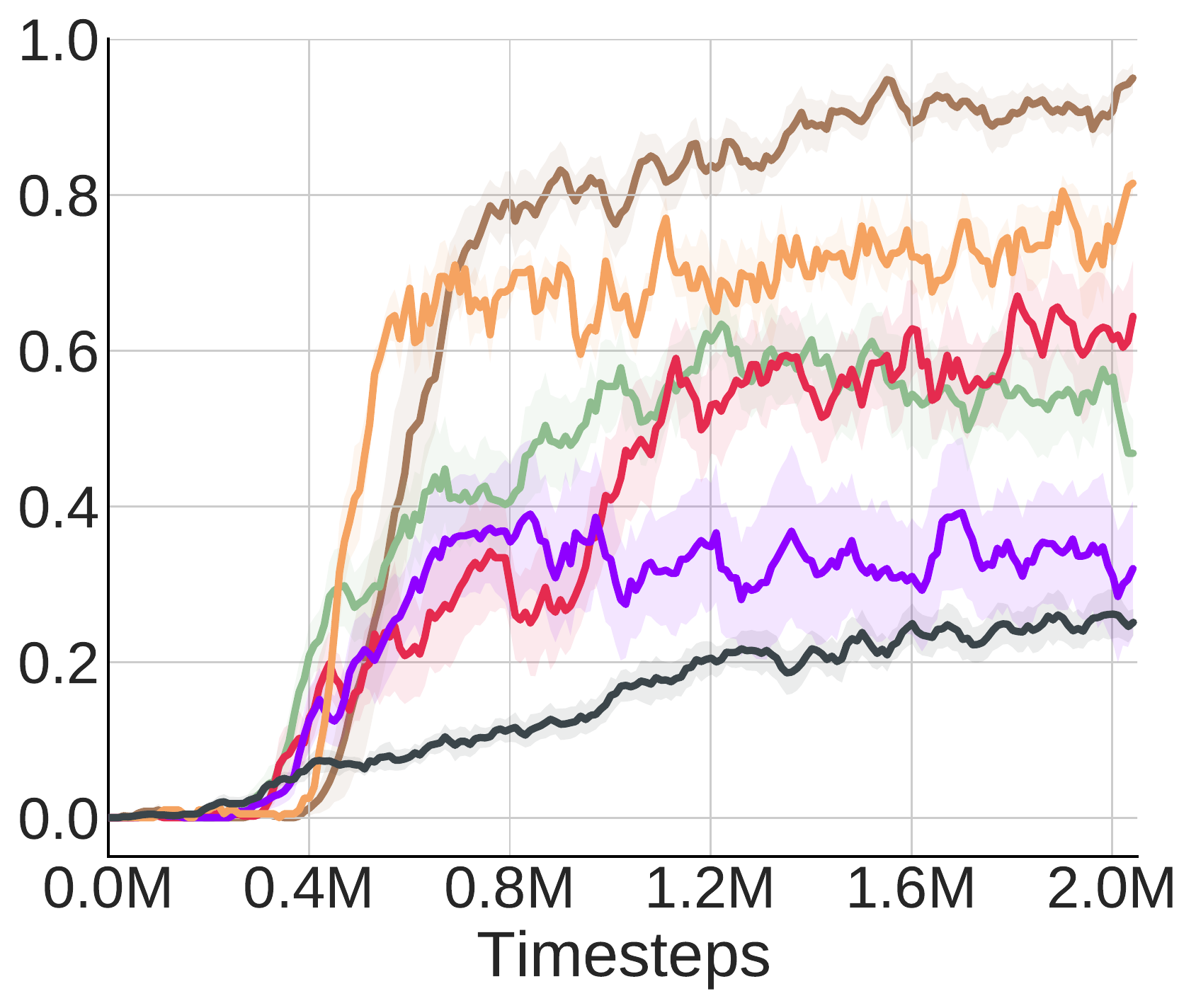}
    }
\subfloat[TJ: slow]{
    \label{tj1}
\includegraphics[width=0.225\textwidth]{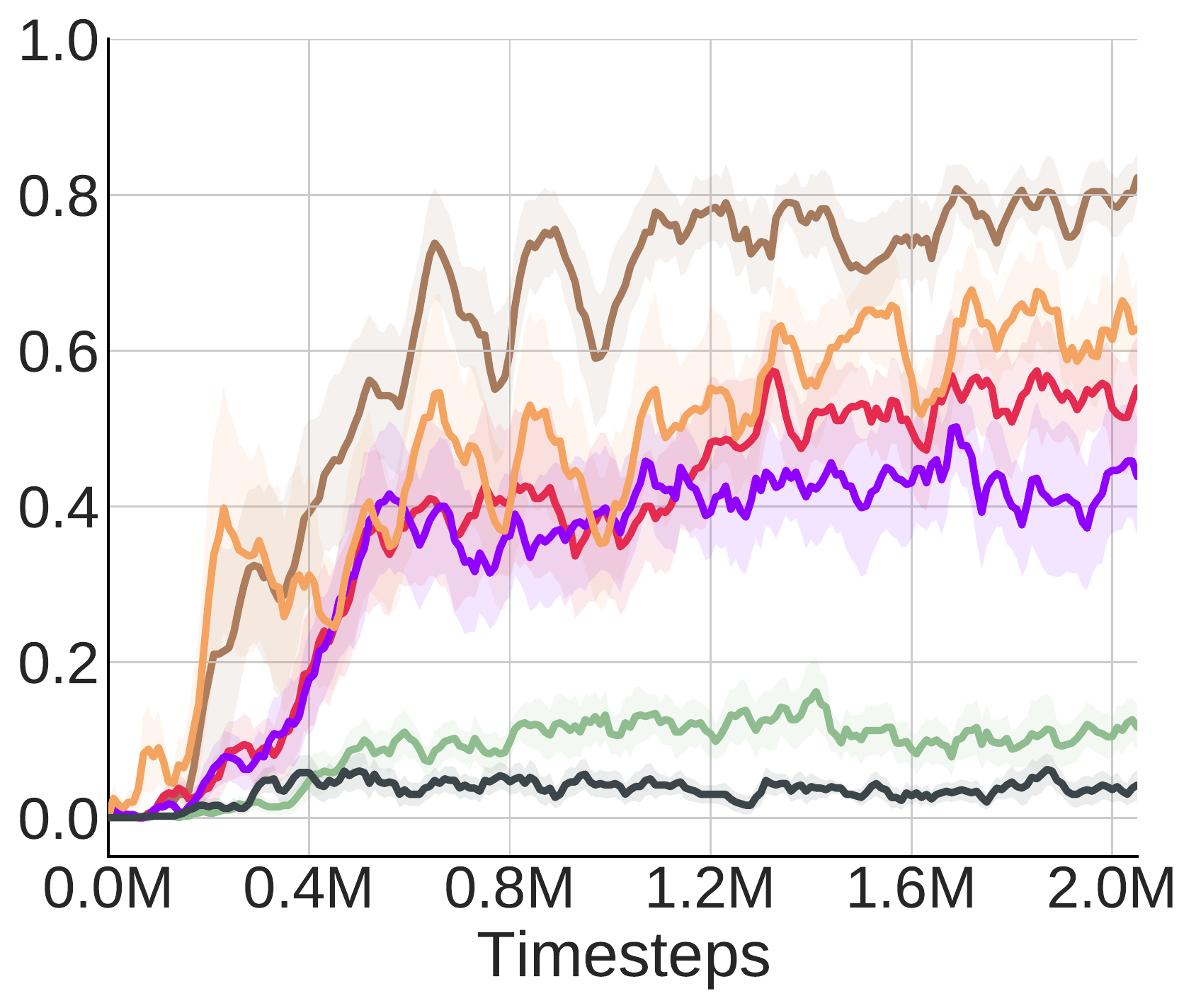}
    }
 \subfloat[SMAC: 1o10b\_vs\_1r]{
    \label{smac1}
\includegraphics[width=0.225\textwidth]{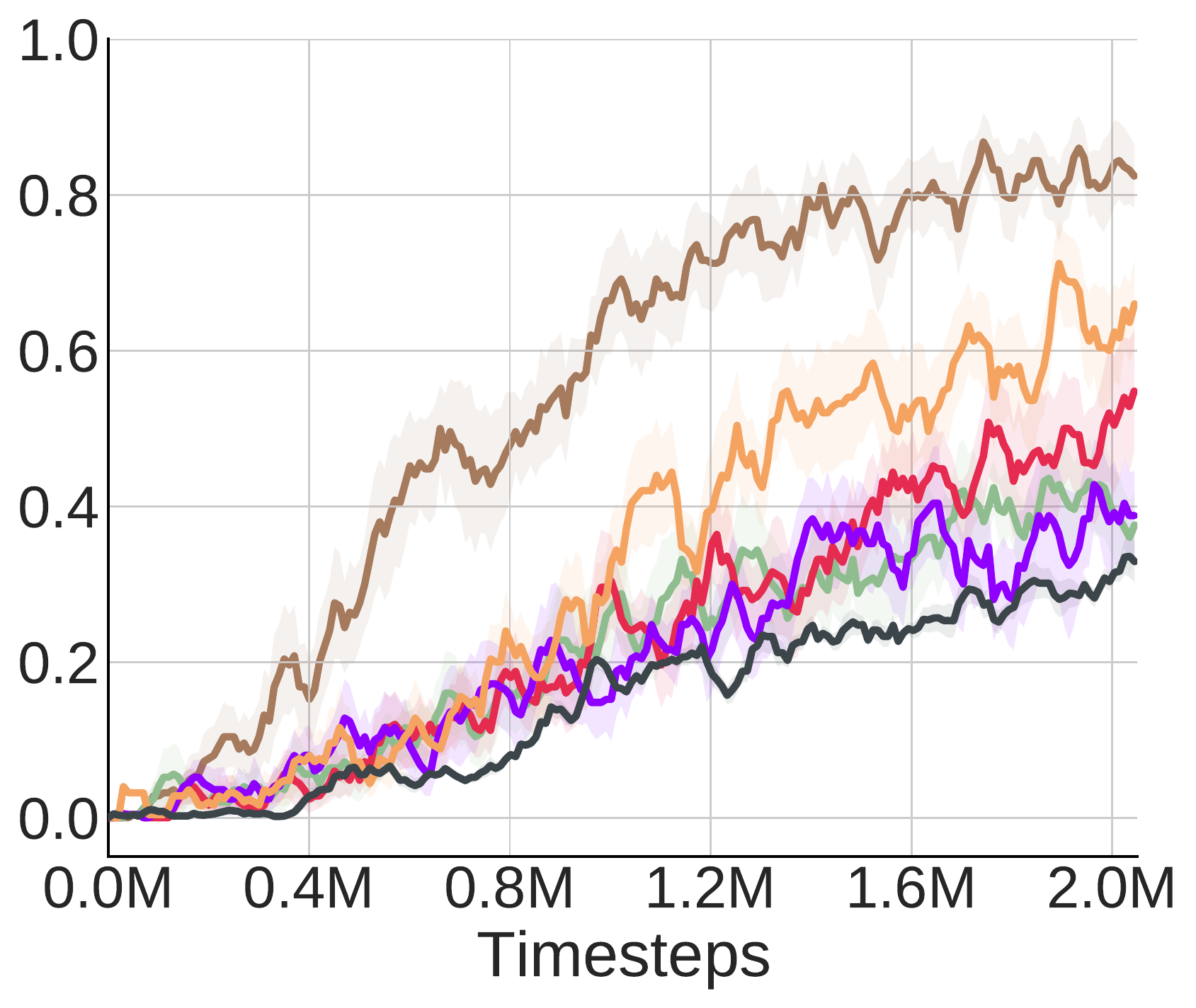}
    }   
    \\
    \subfloat[Hallway: 3x3x4x4]{
    \label{hallway2}
    \includegraphics[width=0.24\textwidth]{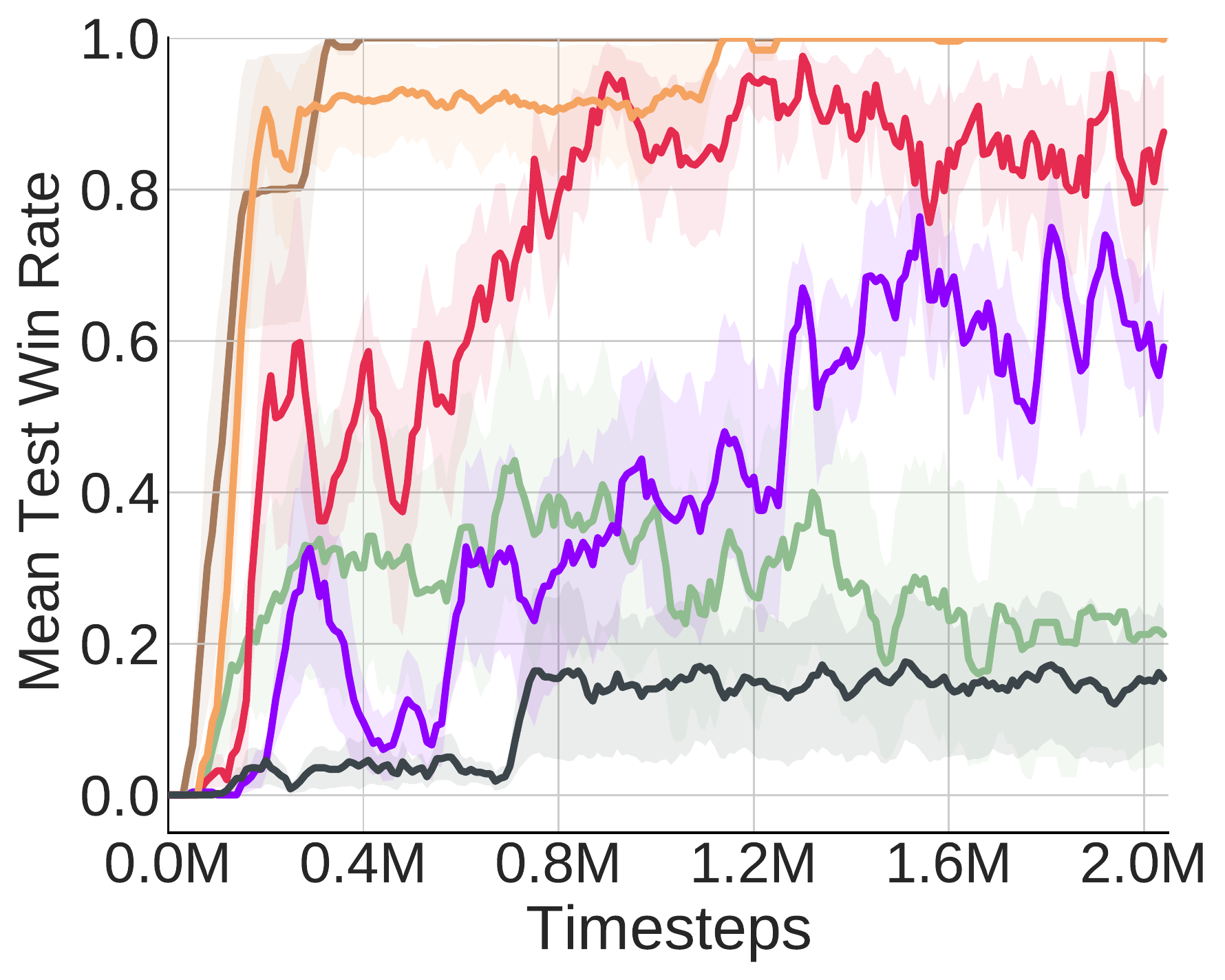}
    } 
    \subfloat[LBF: 4p-1f]{
    \label{lbf2}
    \includegraphics[width=0.225\textwidth]{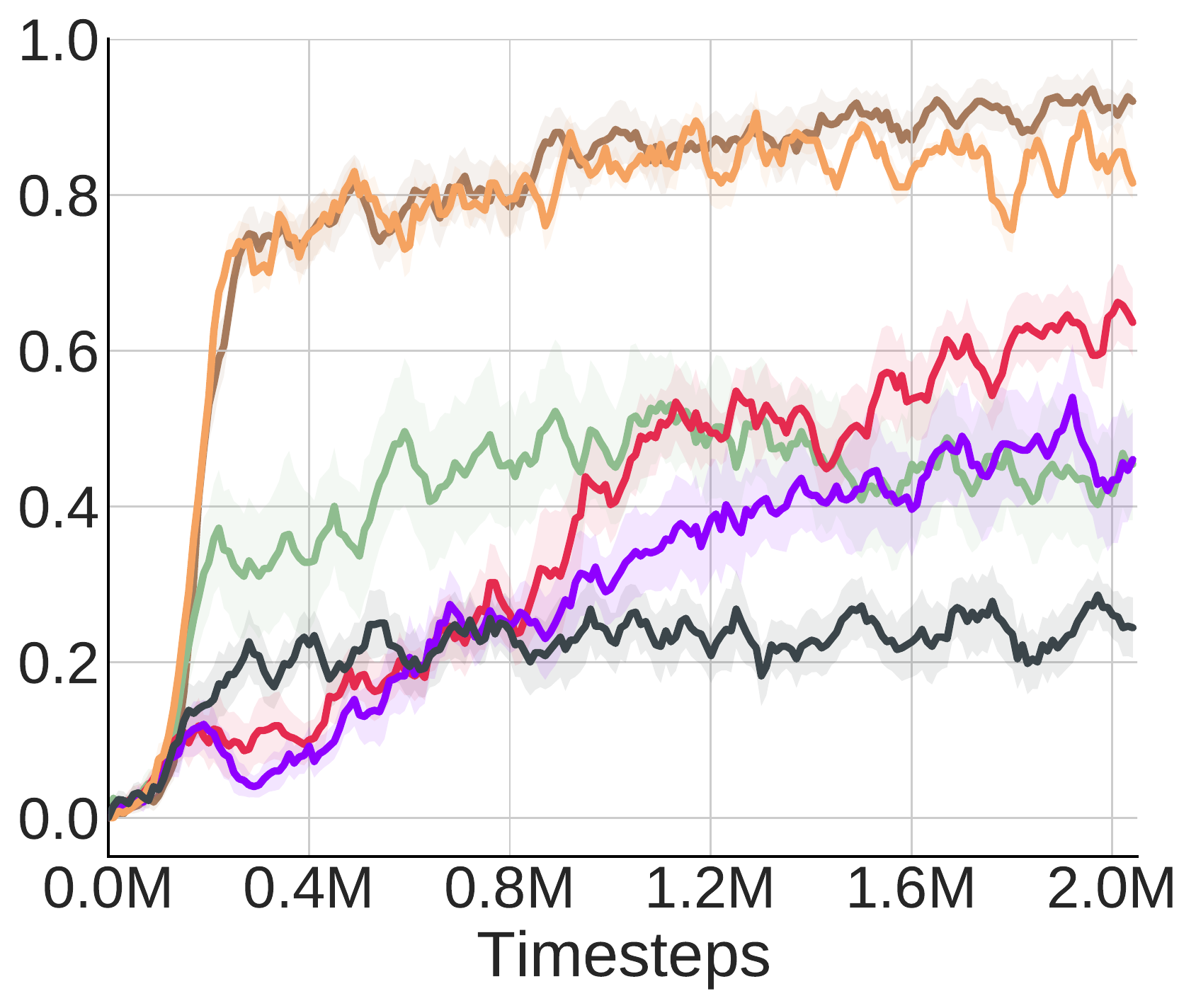}
    }
     \subfloat[TJ: fast]{
    \label{tj2}
    \includegraphics[width=0.225\textwidth]{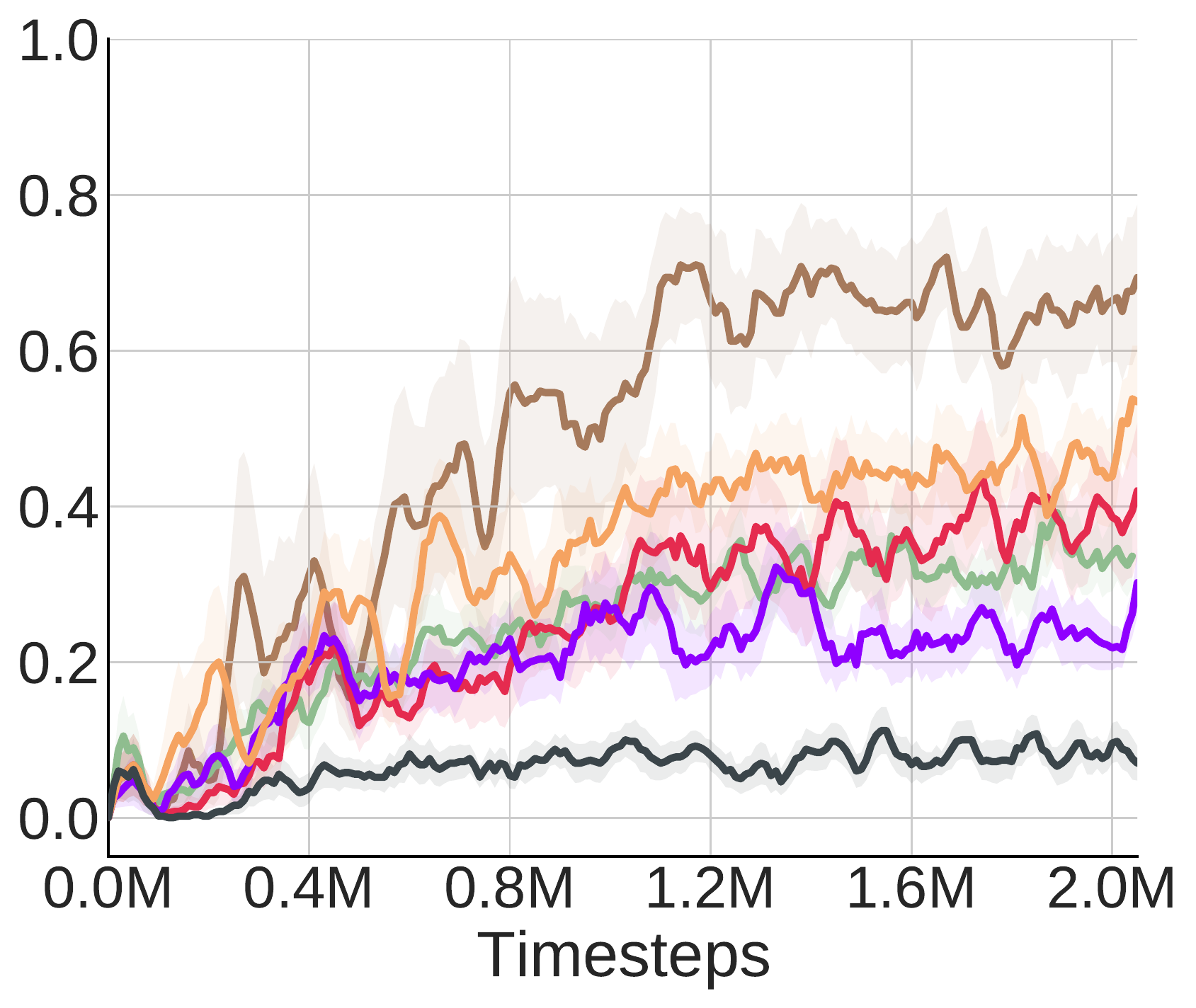}
    } 
   \subfloat[SMAC: 1o2r\_vs\_4r]{
    \label{smac2}
    \includegraphics[width=0.235\textwidth]{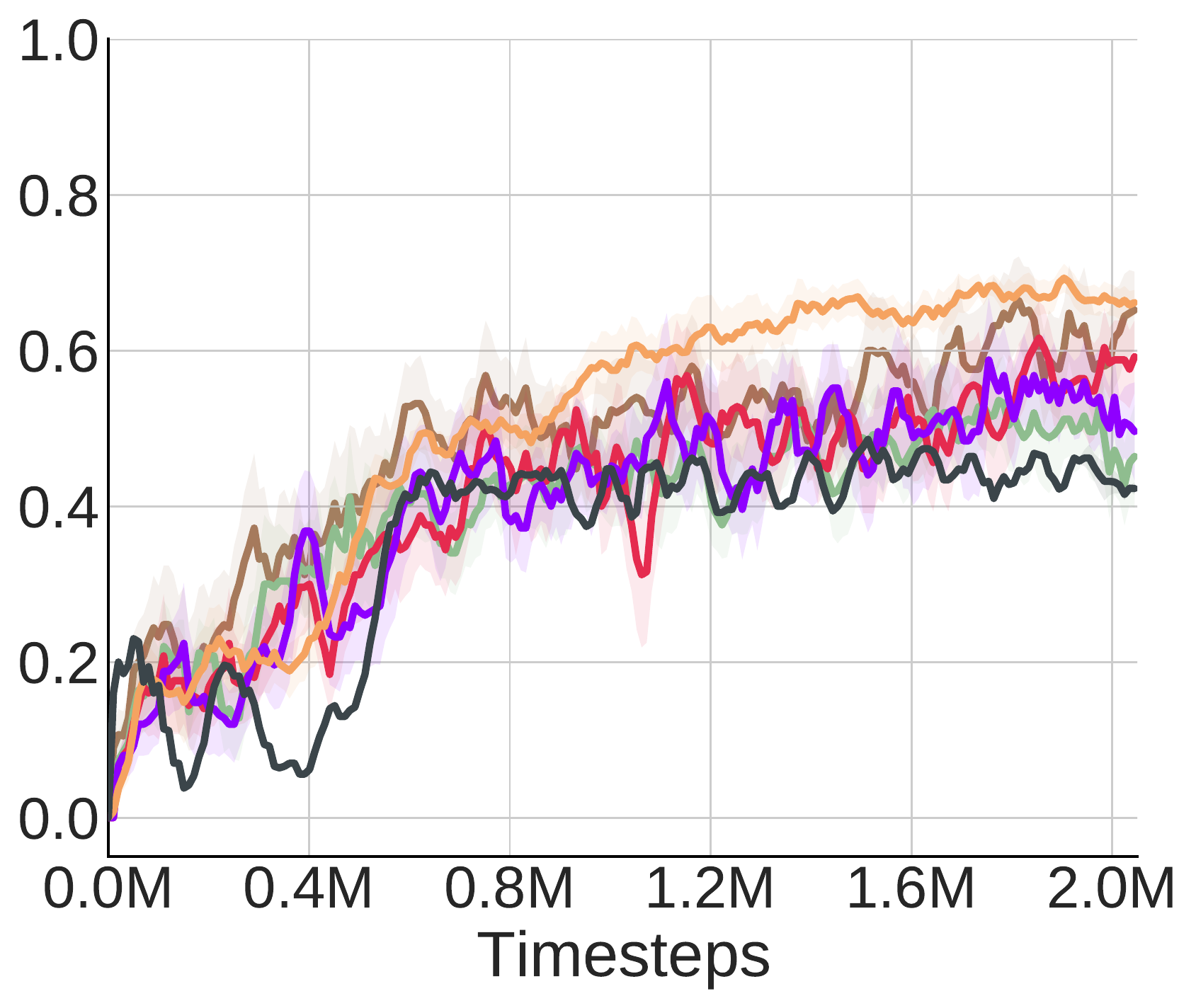}
    } 
    \\
	\subfloat{
    \label{legendsmac}
\includegraphics[width=0.93\textwidth]{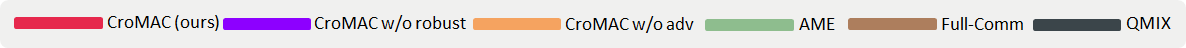}
    }
	\caption{Empirical Results of several algorithms tested in two different perturbation conditions on benchmarks. Note that Full-Comm, CroMAC w/o adv, and QMIX are tested in perturbation-free conditions, while CroMAC, CroMAC w/o robust, and AME suffer from message perturbations when testing. See text for more details. }
	\label{totalperformance}
\end{figure*}
We first compare CroMAC against multiple baselines to investigate the communication robustness on various benchmarks. As shown in Fig.~\ref{totalperformance}, QMIX achieves the most inferior performance in all environments, demonstrating that communication is needed. Full-Comm can solve all the tasks under perturbation-free conditions, showing that these tasks need communication and can be solved by a simple communication mechanism. CroMAC w/o adv, an ablation of CroMAC where testing is conducted under perturbation-free conditions, can achieve comparable coordination ability with Full-Comm, validating the specific design of our CroMAC does not cause much performance degradation for a communication goal. On the contrary, when message perturbations occur during the testing phase, it can be easily found that CroMAC w/o robust, a variant of our proposed method without an efficient robust mechanism, suffers from severe performance degradation compared with Full-Comm and CroMAC w/o adv. However, CroMAC exhibits higher robustness than others, and it surprises us that AME also suffers from severe performance degradation under perturbations, which means an unreasonable constraint for robustness training cannot be applied in complex and severe message perturbation conditions.
\begin{figure*}
  \centering
  \includegraphics[width=1\textwidth]{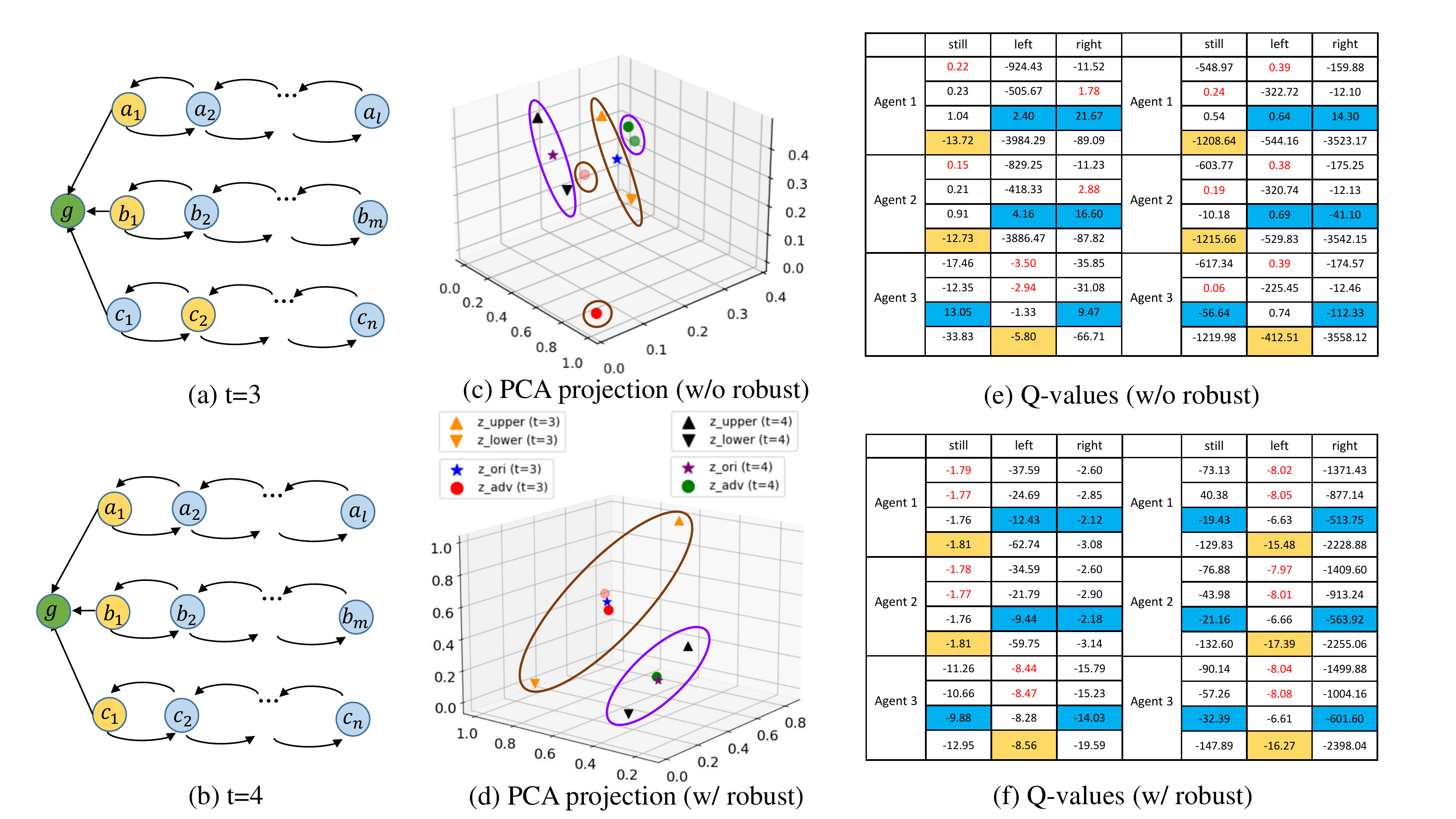}
  \caption{Visualization results. We take $t=3$ and $4$ in Hallway as shown in (a) and (b),  where Agent 1 and Agent 2 stand one step from the goal while Agent 3 needs to take two steps to reach the goal. (c) and (d) show the PCA projection~\cite{tipping1999probabilistic} of the message representation $z_{\rm msg}$ for (a) and (b), with  $\bullet$ and $\star$ represent $z_{\rm msg}$ with and  without perturbations, respectively. Note that $z_{\rm msg}$ is the same for agents without perturbations, and some $\bullet$ are darker because multiple ones overlap together. $\blacktriangle,\blacktriangledown$ represents the upper and lower bounds of $z_{\rm msg}$, note that ellipses of the same color represent the same time step. (e) and (f) display the $Q$-values (multiplied by 100 for viewing) of each agent accordingly, where the first row means the original $Q$-values of all actions, while the second row refers to them under perturbations with red fonts representing the selected actions in corresponding cases. The third and fourth rows show the upper and lower bounds of $Q$-values under $\mathbb{\epsilon}$-perturbation, where yellow squares are the lower bounds of $Q$-values over best actions while blue squares are the upper bounds of $Q$-values over other actions. }
  \label{vision}
\end{figure*}

Furthermore, we conduct experiments on task Hallway  to investigate how CroMAC learns a robust communication policy. As shown in Fig.~\ref{vision}, three agents coordinate to reach the goal. When suffering from perturbations, the message representation learned by methods without a robust mechanism will go out of the upper and lower bounds, leading to an unpredictable input for the local policy. Consequently, the message perturbation influences the action selection of each agent.
Take Agent 1 in Fig.~\ref{vision}(e) as an example. It should keep still with Agent 2 at $t=3$ to wait for Agent 3 to 
go left together for success. However, when suffering from perturbations, the message representation jumps out of the normal range. It unexpectedly goes right, as the according $Q$-value $1.78$ is dominant to others ($0.23$ for still and $-505.67$ for left). On the contrary, with our robustness scheme, the message representations can be bounded in a reasonable range, leading to a robust action selection compared with the perturbation-free setting, as shown in Fig.~\ref{vision}(f). The whole process shows our approach can obtain message certification when any perturbations happen.



\subsection{Robustness Under Various Perturbations } \label{generalization}
As this study considers a setting where the number of attacks is fixed during the training phase, we evaluate here the generalization ability when altering the perturbation budget and encountering different perturbation methods in the testing phase. Specifically, we conduct experiments on each benchmark with the same structures and hyperparameters as Sec.~\ref{resultandanalysis} during training. As shown in Tab.~\ref{messagepruing}, we 
consider eight communication situations, where “Natural” means no perturbation exist, FGSM is the training condition of the comparable approaches, and others like PGD are other conditions, details can be seen in~\ref{detailedimplement}. We can find that AME can achieve comparable or even superiority over CroMAC in the Natural setting without message perturbations and also maintain competitiveness when suffering from random perturbations, showing that AME possesses robustness for simple message perturbations. However, AME sustains a drastic performance degradation when we alter the perturbation budget like FGSM (4) or other perturbation models like PGD in the Hallway environment. On the other hand, our CroMAC achieves high superiority over AME in most environments under different perturbations, demonstrating its high generalization ability when encountering different perturbation budgets and perturbation methods.



\begin{table*} 
  \centering
  \caption{Average test win rates for CroMAC and AME under different message perturbation conditions. The results are averaged from 1000 test episodes among 5 random seeds, where FGSM (n) refers to methods under FGSM attack with different budgets. The details of each perturbation methods are shown in \ref{detailedimplement}. }
  \setlength{\tabcolsep}{0.5mm}{
\resizebox{1\textwidth}{!}{
\begin{tabular}{|c|c|c|c|c|c|c|c|c|c|c|}
\midrule
Environment & Method & Natural & Random & PGD & FGSM (1)& FGSM (2)& FGSM & FGSM (3) & FGSM (4)\\
\toprule

\multirow{2}{*}
{\begin{tabular}[c]{@{}c}Hallway\\4x5x6\end{tabular}} & \multicolumn{1}{c|}{CroMAC}& \multicolumn{1}{c|}{0.93$\pm$0.06} &\multicolumn{1}{c|}{0.91$\pm$0.11} &\multicolumn{1}{c|}{\textbf{0.92$\pm$0.13}} 
& \multicolumn{1}{c|}{\textbf{0.97$\pm$0.03}} &\multicolumn{1}{c|}{\textbf{0.86$\pm$0.04}} &\multicolumn{1}{c|}{\textbf{0.91$\pm$0.10}} 
&\multicolumn{1}{c|}{0.60$\pm$0.31} &\multicolumn{1}{c|}{\textbf{0.66$\pm$0.39}} 
\\
& \begin{tabular}[c]{@{}c@{}}AME \end{tabular} & 0.98$\pm$0.01 & 0.93$\pm$0.04 & 0.43$\pm$0.20& 0.66$\pm$0.34 & 0.61$\pm$0.34 & 0.62$\pm$0.31& 0.36$\pm$0.10& 0.10$\pm$0.20\\ 
& \begin{tabular}[c]{@{}c@{}}REC \end{tabular} & \textbf{1.00$\pm$0.00} & \textbf{0.95$\pm$0.08} & 0.90$\pm$0.20& 0.96$\pm$0.06 & 0.62$\pm$0.38 & 0.82$\pm$0.23& \textbf{0.68$\pm$0.40}& 0.41$\pm$0.43\\ 
\midrule

\multirow{2}{*}{\begin{tabular}[c]{@{}c}LBF\\3p-1f\end{tabular}} & \multicolumn{1}{c|}{CroMAC}& \multicolumn{1}{c|}{0.71$\pm$0.05} &\multicolumn{1}{c|}{0.72$\pm$0.03} &\multicolumn{1}{c|}{\textbf{0.61$\pm$0.09}} 
& \multicolumn{1}{c|}{\textbf{0.71$\pm$0.07}} &\multicolumn{1}{c|}{\textbf{0.67$\pm$0.09}} &\multicolumn{1}{c|}{\textbf{0.64$\pm$0.13 }} 
&\multicolumn{1}{c|}{\textbf{0.43$\pm$0.15}} &\multicolumn{1}{c|}{\textbf{0.30$\pm$0.08}} 
\\
& \begin{tabular}[c]{@{}c@{}}AME \end{tabular} & \textbf{0.77$\pm$0.04} & 0.72$\pm$0.04 & 0.58$\pm$0.11& 0.63$\pm$0.09 & 0.56$\pm$0.02 & 0.47$\pm$0.03& 0.36$\pm$0.10& 0.29$\pm$0.04\\ 
\midrule

\multirow{2}{*}{\begin{tabular}[c]{@{}c}TJ\\slow\end{tabular}} & \multicolumn{1}{c|}{CroMAC}
&\multicolumn{1}{c|}{\textbf{0.31$\pm$0.07}} &\multicolumn{1}{c|}{\textbf{0.46$\pm$0.20}} &\multicolumn{1}{c|}{\textbf{0.31$\pm$0.23}} 
&\multicolumn{1}{c|}{\textbf{0.29$\pm$0.12}} &\multicolumn{1}{c|}{\textbf{0.31$\pm$0.09}} &\multicolumn{1}{c|}{\textbf{0.37$\pm$0.14}} 
&\multicolumn{1}{c|}{\textbf{0.42$\pm$0.16}} &\multicolumn{1}{c|}{\textbf{0.32$\pm$0.18}} 
\\
& \begin{tabular}[c]{@{}c@{}}AME \end{tabular} & {0.12$\pm$0.07} & 0.13$\pm$0.03 & 0.15$\pm$0.06& 0.13$\pm$0.06 & 0.13$\pm$0.06 & 0.12$\pm$0.06& 0.08$\pm$0.02& 0.13$\pm$0.06\\ 
\midrule

\multirow{2}{*}{\begin{tabular}[c]{@{}c}SMAC\\1o10b\_vs\_1r\end{tabular}} & \multicolumn{1}{c|}{CroMAC}& \multicolumn{1}{c|}{\textbf{0.65$\pm$0.10 }} &\multicolumn{1}{c|}{\textbf{0.64$\pm$0.12}} &\multicolumn{1}{c|}{\textbf{0.53$\pm$0.07}} 
& \multicolumn{1}{c|}{\textbf{0.56$\pm$0.04}} &\multicolumn{1}{c|}{\textbf{0.52$\pm$0.18}} &\multicolumn{1}{c|}{\textbf{0.59$\pm$0.08}} 
&\multicolumn{1}{c|}{{0.41$\pm$0.14}} &\multicolumn{1}{c|}{{0.34$\pm$0.03}} 
\\
& \begin{tabular}[c]{@{}c@{}}AME \end{tabular} & 0.38$\pm$0.12 & 0.52$\pm$0.24 & 0.51$\pm$0.17& 0.44$\pm$0.13 & 0.45$\pm$0.20 & 0.38$\pm$0.02& \textbf{0.44$\pm$0.19}& \textbf{0.43$\pm$0.07}\\ 
\midrule

\end{tabular}}}
  
  \label{messagepruing}
\end{table*}

\begin{figure*}
\centering
\subfloat[LBF: 3p-1f]{
\label{methodintegrative}
\includegraphics[width=0.45\textwidth]{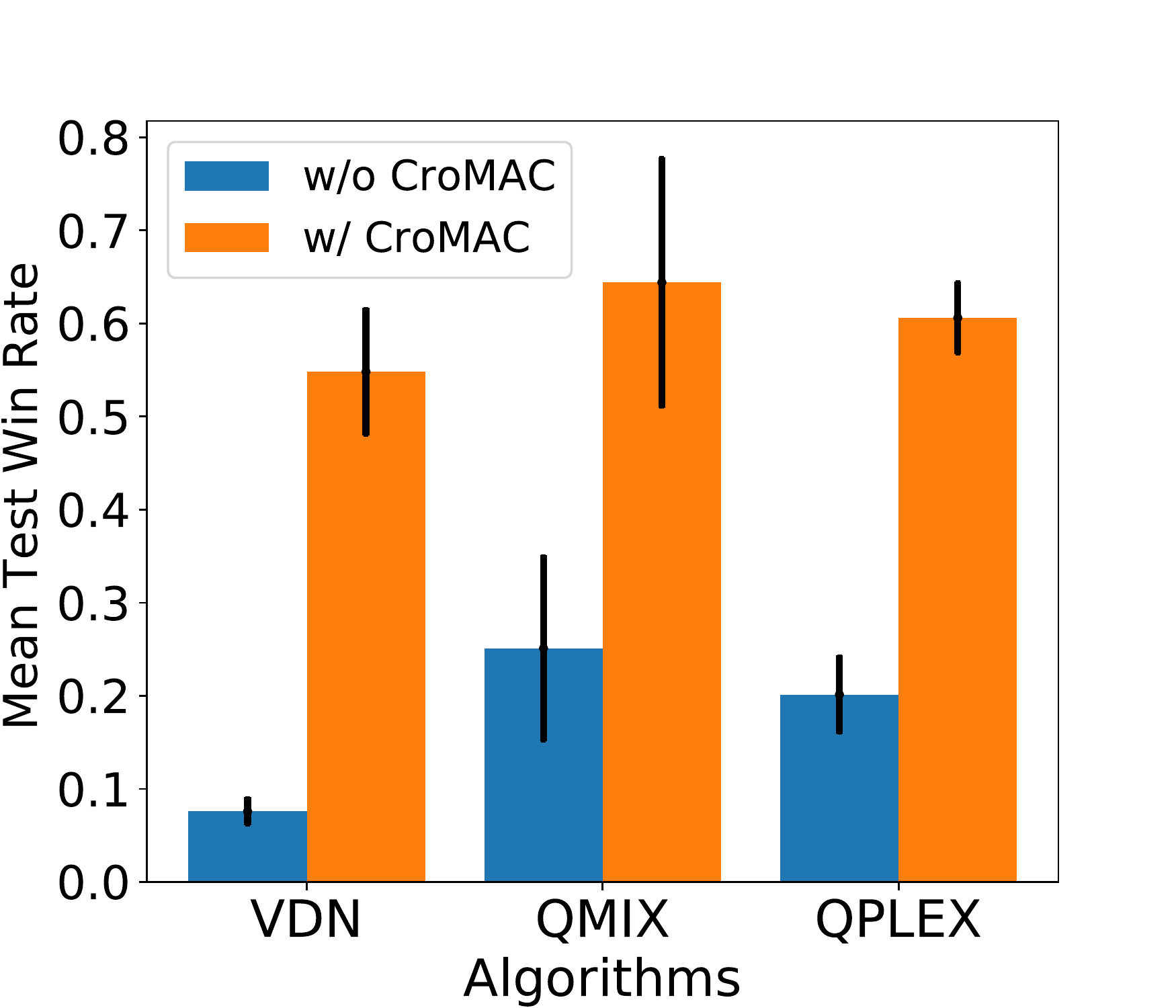}
}
\subfloat[SMAC: 1o10b\_vs\_1r]{
\label{Integrative2}
\includegraphics[width=0.45\textwidth]{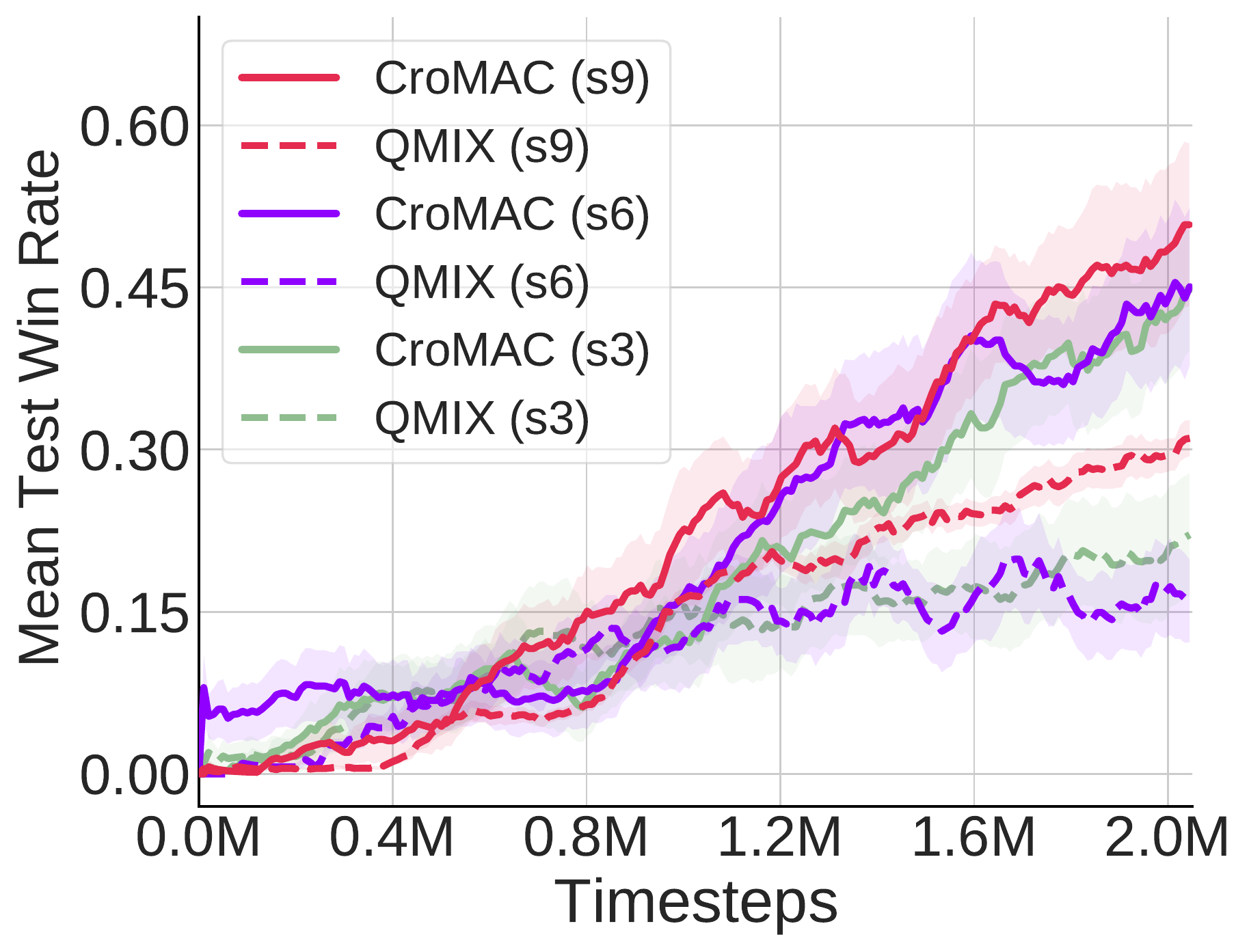}
 }
\caption{(a) Average test success rates of CroMAC implemented with different value based MARL methods. (b) Performance comparison with varying sights, where $sn$ means the sight range is $n$ and the default sight range is 9.}
\end{figure*}

\subsection{Generality and Parameter Sensitivity} \label{sensitive}
CroMAC is a robust communication training paradigm and is agnostic to specific value decomposition MARL methods and any sight conditions. Here we treat it as a plug-in module and integrate it with existing MARL value decomposition methods like VDN~\cite{vdn},   QMIX~\cite{qmix}, and QPLEX~\cite{qplex}. As shown in Fig.~\ref{methodintegrative}, when integrating with CroMAC, the performance of the baselines vastly improves on scenario LBF:3p-1f with message perturbations, indicating that the proposed training paradigm can significantly enhance robustness for various MARL methods. It is worth mentioning that QPLEX shows instability when adding robust loss and we choose the best result in the training process for comparison. Furthermore, when we alter the agents’ sight range in the SMAC map 1o10b\_vs\_1r, the results shown in Fig.~\ref{Integrative2} also demonstrate that CroMAC can improve the coordination 
robustness in different sight conditions for QMIX with communication, showing its high generality under different communication scenarios.  

\begin{figure}
\centering
\subfloat[Sensitivity of $\kappa$]{
\label{kappa}
\includegraphics[width=0.33\textwidth]{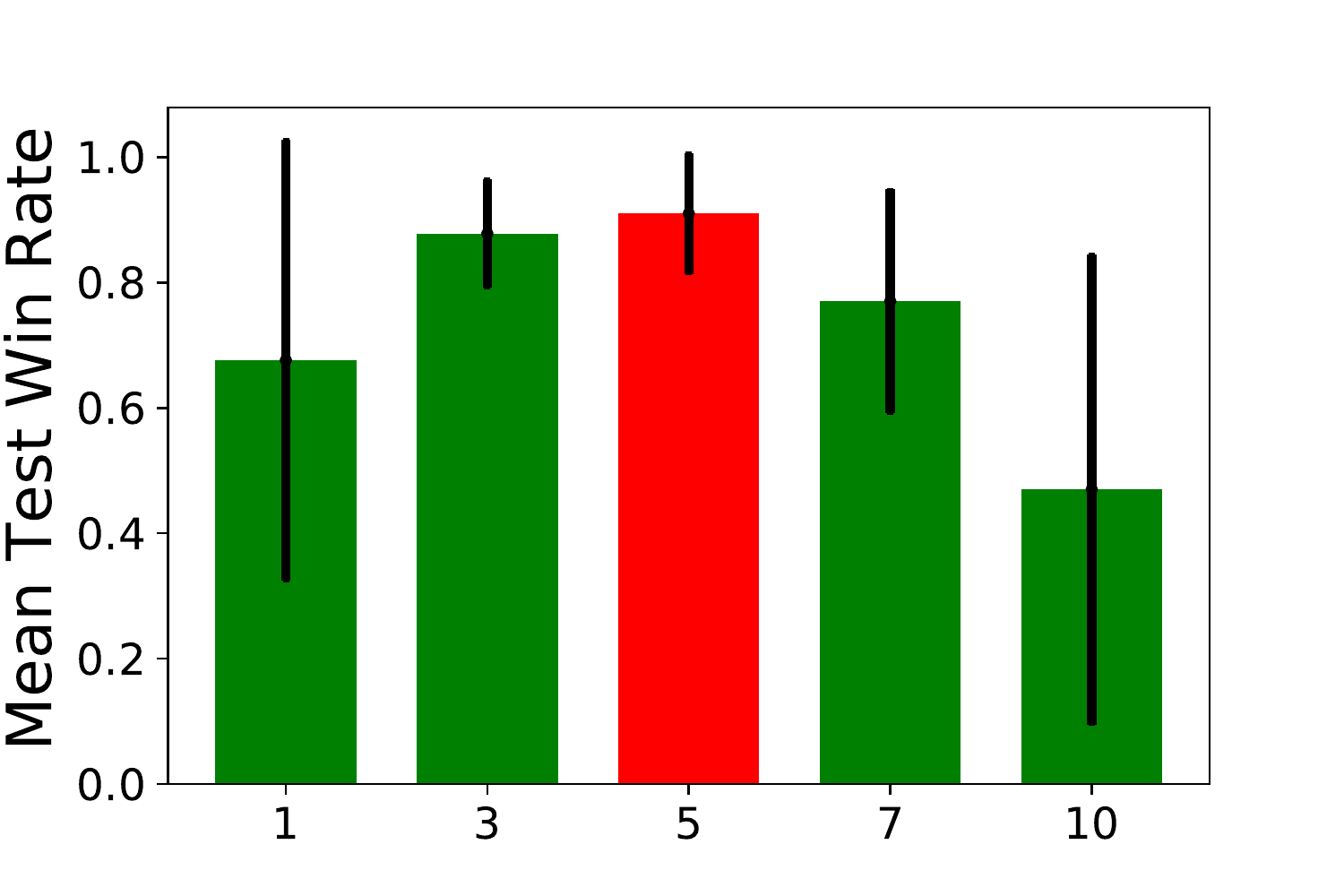}
} 
\subfloat[Sensitivity of $C\_MAX$]{
\label{cminmax}
\includegraphics[width=0.33\textwidth]{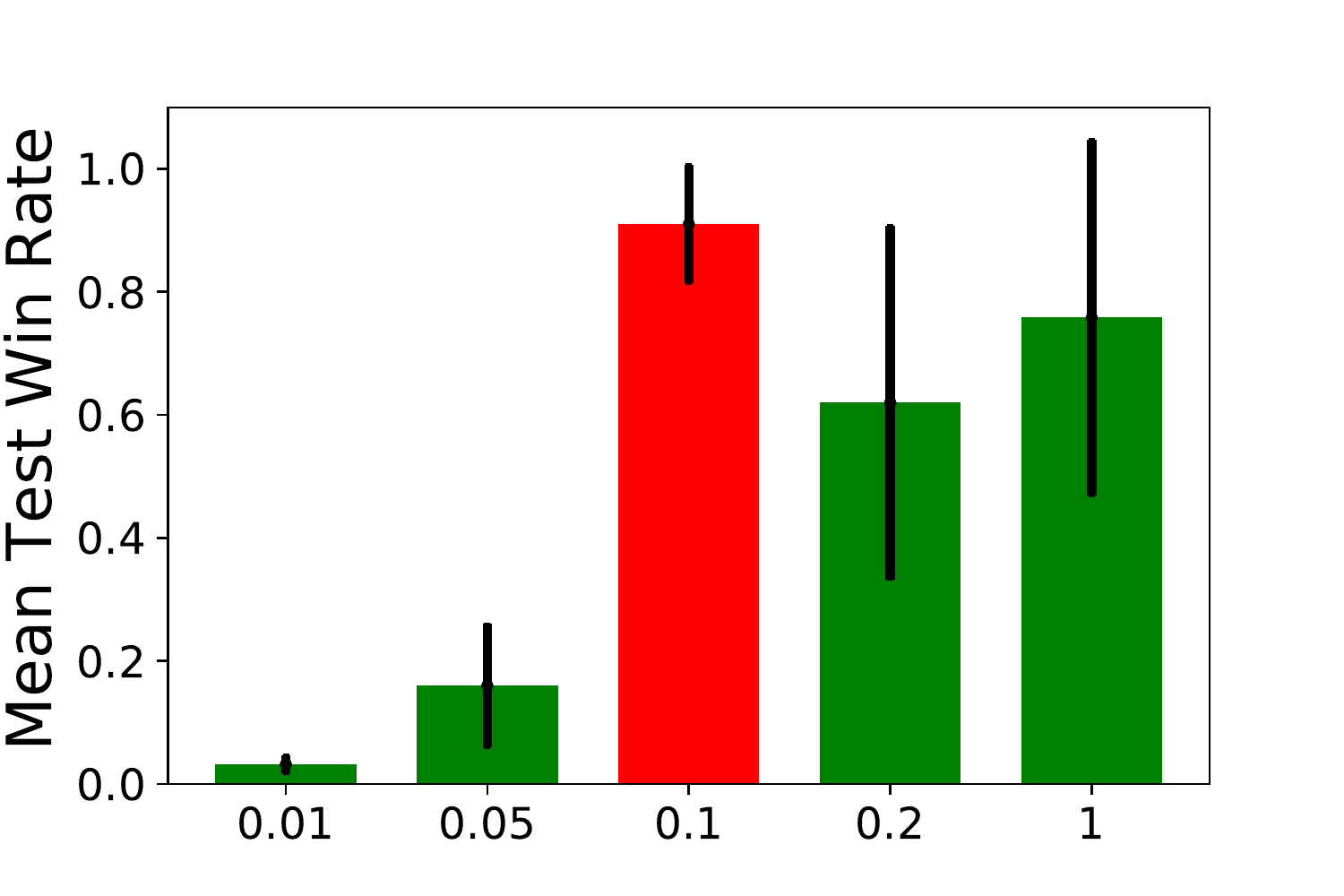}
} 
\\
\subfloat[Sensitivity of $\alpha_1$]{
\label{alpha1}
\includegraphics[width=0.33\textwidth]{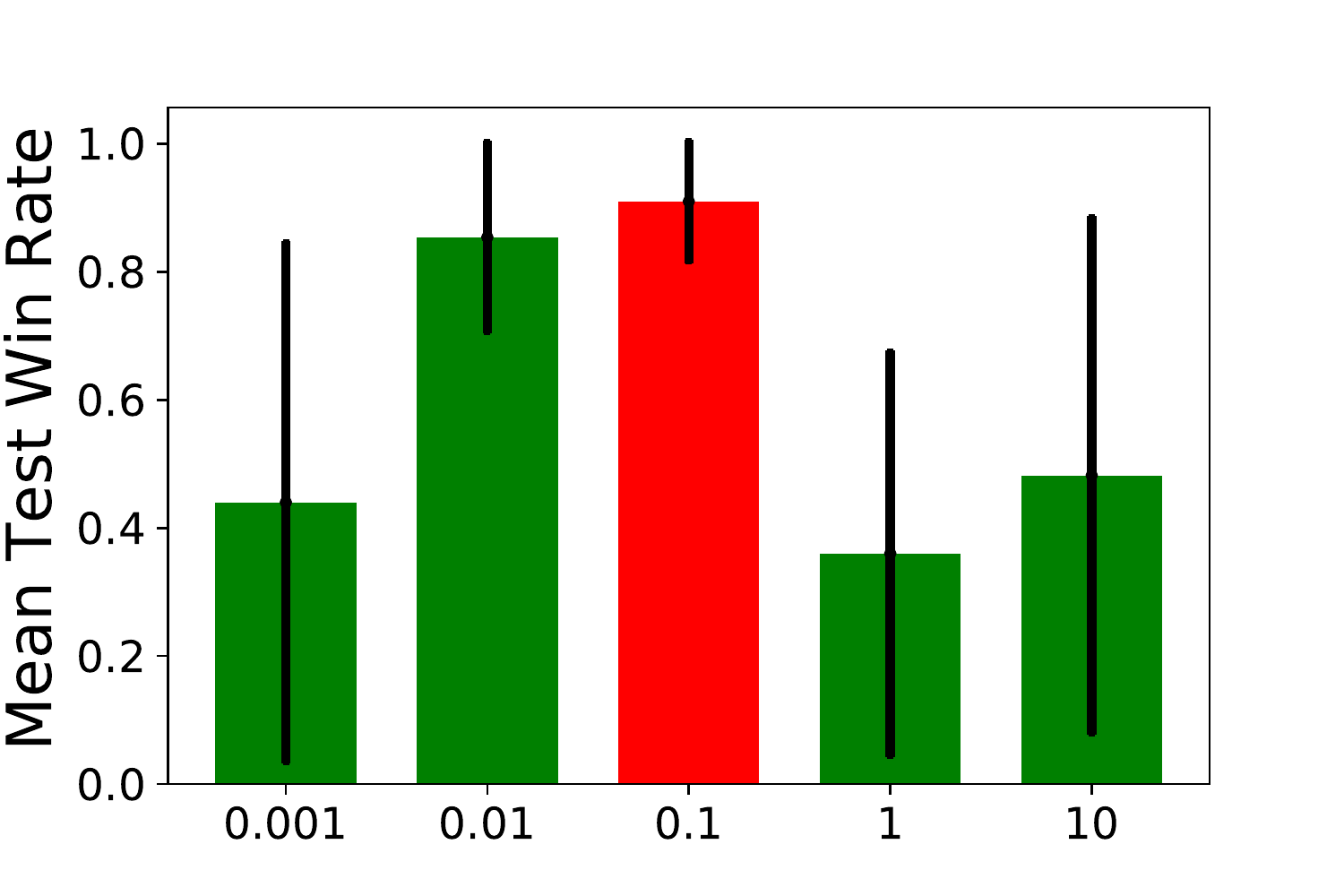}
}
\subfloat[Sensitivity of $\alpha_2$]{
\label{alpha2}
\includegraphics[width=0.33\textwidth]{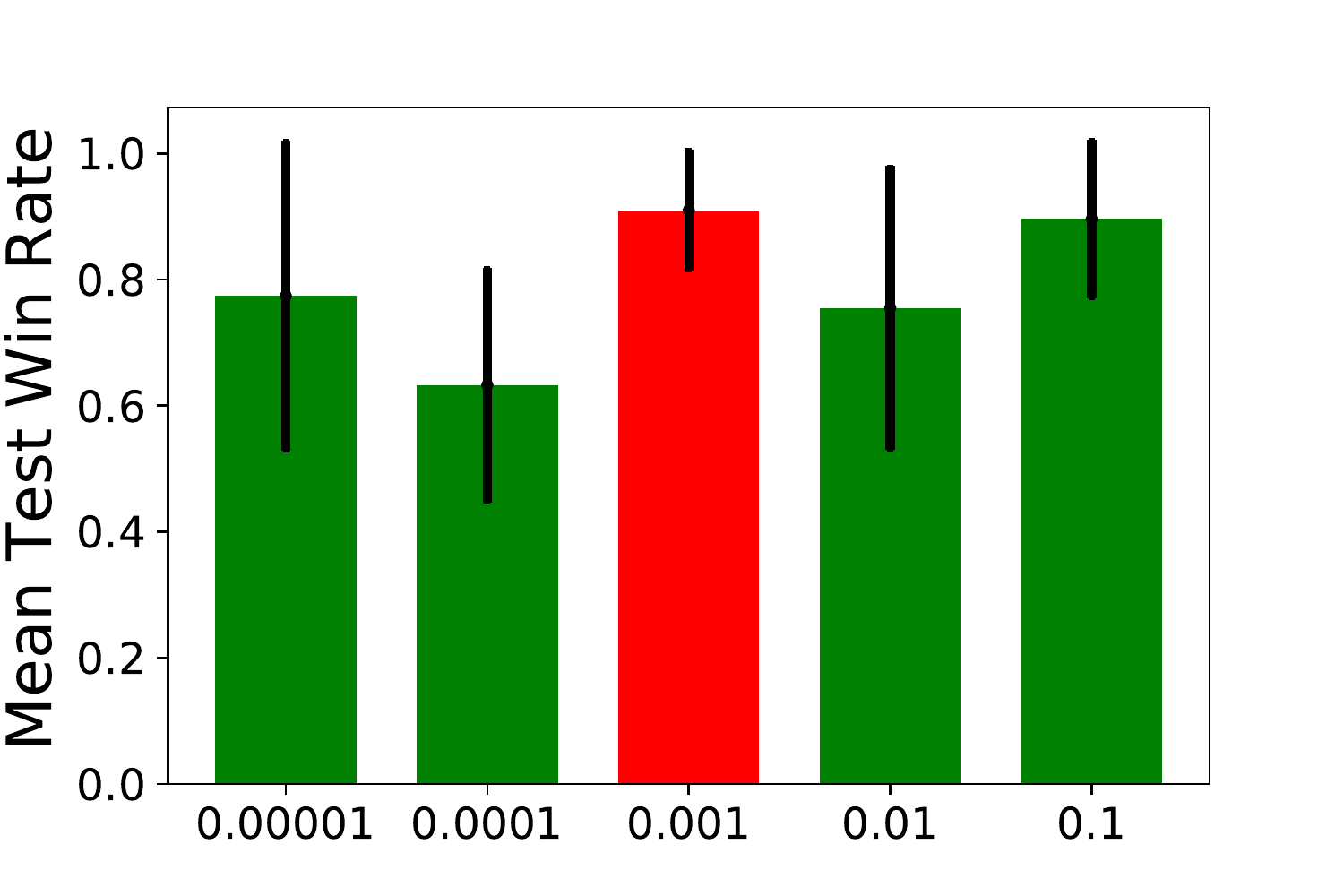}
}
\subfloat[Sensitivity of $\alpha_3$]{
\label{alpha3}
\includegraphics[width=0.33\textwidth]{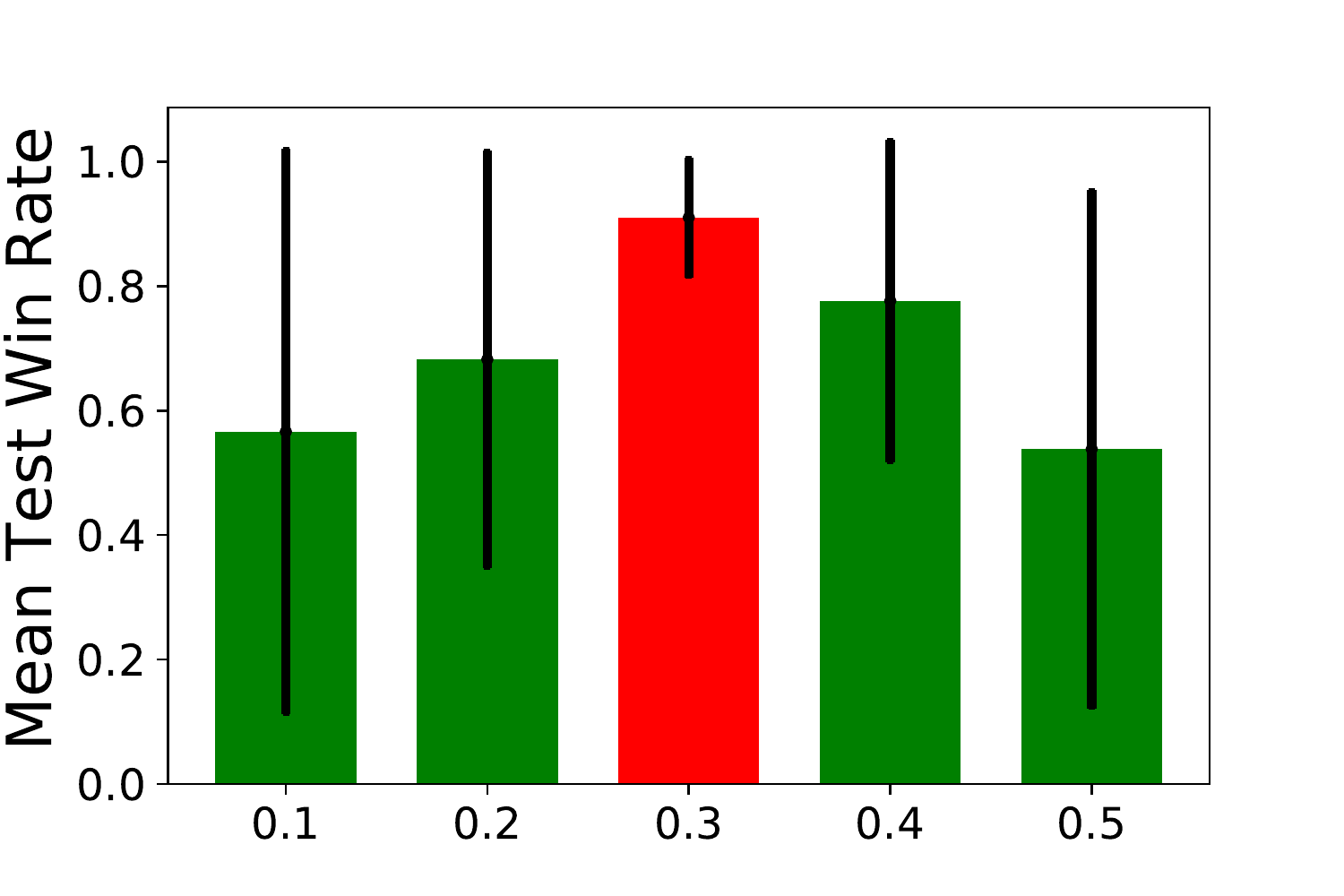}
}

\caption{Sensitivity of hyperparameters used in this paper.}
\end{figure}

As CroMAC includes multiple hyperparameters, here we conduct experiments on scenario Hallway: 4x5x6 to investigate how each one influences the robustness. Where $\kappa$ controls the attack strength added to the latent state space. If it is too small, we cannot guarantee good robustness in the testing phase, and if it is too large, the policy may be too smooth and not optimal anymore. As shown in Fig.~\ref{kappa}, we can find that $\kappa=3$ is
the best choice in this scenario. Furthermore, the value range of the network weight $W$ is used to get an approximate bound of the integration error. Fig.~\ref{cminmax} shows that $C\_MAX=0.1$ performs best. We can find the most appropriate parameters for other scenarios in the same way. More details for other scenarios are shown in App.~E. Besides, as there are multiple hyperparameters of each loss function, we show how each adjustable hyperparameter named $\alpha_1, \alpha_2$, and $\alpha_3$ influence the robustness of CroMAC, we continue to conduct experiments on the task Hallway: 4x5x6 to investigate how each hyperparameter $\alpha$ influences the robustness. As shown in Fig.~\ref{alpha1}-\ref{alpha3}, we can find that when the parameter is slightly larger or smaller, the performance may suffer corresponding degradation and the stability will also decline.


\section{Conclusion and Future Work}
Considering the great significance of robustness for real-world policy deployment and the enormous potential of MARL,
 this paper takes a further step towards robustness in MARL communication. We first model the multi-agent communication as a multi-view problem and apply a multi-view variational autoencoder that uses a product-of-experts inference network to obtain a joint message representation from the received messages, then a certificate guarantee between the joint message representation and each received message is obtained via interval bound propagation. For the optimization phase, we first encode the state into a latent space,  
and do perturbations in this space to get a certificate state representation. Then the learned joint message representation is used to approximate the certificate state representation. Extensive experimental results from multiple aspects demonstrate the efficiency of the proposed method. In terms of possible future work, as we learn the communication policy online, how we can learn a robust communication policy in offline MARL is challenging but of great value.

\section*{Acknowledgements}

This work is partly supported by the National Key Research
and Development Program of China (2020AAA0107200), the National Science
Foundation of China (61921006, 61876119, 62276126), the Natural
Science Foundation of Jiangsu (BK20221442), and the program B for Outstanding PhD candidate of Nanjing University. We thank Ziqian Zhang and Fuxiang Zhang for their useful suggestions.


\bibliographystyle{ACM-Reference-Format}
\bibliography{SCIS}




 \newpage
\begin{appendix}
\section{Product of a Finite Number of Gaussians} \label{poeprof}
Suppose we have $N$ Gaussian experts with means $\mu_{i1},\mu_{i2},\cdots,\mu_{iN}$ and variances $\sigma_{i1}^2,\sigma_{i2}^2,\cdots,\sigma_{iN}^2$, the product distribution is still Gaussian with mean $\mu_i$ and variance $\sigma_i^2$:
\begin{equation} \label{eqn.qwe}
\begin{aligned}
    \mu_i &= \Big(\frac{\mu_{i1}}{\sigma_{i1}^2}+\frac{\mu_{i2}}{\sigma_{i2}^2}+\cdots+\frac{\mu_{iN}}{\sigma_{iN}^2}\Big)\sigma_i^2, \\
    \frac{1}{\sigma_i^2} &= \frac{1}{\sigma_{i1}^2} + \frac{1}{\sigma_{i2}^2} + \cdots + \frac{1}{\sigma_{iN}^2}.
\end{aligned}
\end{equation}
It can be proved by induction.

\begin{proof}

We want to prove Eqn.~\ref{eqn.qwe} is true for all $N\ge 2$.

\begin{itemize}
    \item Base case: Suppose $N=2$ and $p_1(x)=\mathcal{N}(x|\mu_1,\sigma_1),p_2(x)=\mathcal{N}(x|\mu_2,\sigma_2)$, then\begin{align}
        p_1(x)p_2(x) &= \frac{1}{\sqrt{2\pi}\sigma_1}\exp{\left(-\frac{\left(x-\mu_1\right)^2}{2\sigma_1^2}\right)}\cdot\frac{1}{\sqrt{2\pi}\sigma_2}\exp{\left(-\frac{\left(x-\mu_2\right)^2}{2\sigma_2^2}\right)}\nonumber \\
        &= \frac{1}{2\pi\sigma_1\sigma_2}\exp{\left(-\left(\frac{\left(x-\mu_1\right)^2}{2\sigma_1^2}+\frac{(x-\mu_2)^2}{2\sigma_2^2}\right)\right)} \nonumber \\
        &= \frac{1}{2\pi\sigma_1\sigma_2}\exp{\left(-\frac{x^2-2\frac{\mu_1\sigma_2^2+\mu_2\sigma_1^2}{\sigma_1^2+\sigma_2^2}x+\frac{\mu_1^2\sigma_2^2+\mu_2^2\sigma_1^2}{\sigma_1^2+\sigma_2^2}}{2\frac{\sigma_1^2\sigma_2^2}{\sigma_1^2+\sigma_2^2}}\right)}\nonumber \\
        &= \frac{1}{2\pi\sigma_1\sigma_2}\exp{\left(-\frac{\left(x-\frac{\mu_1\sigma_2^2+\mu_2\sigma_1^2}{\sigma_1^2+\sigma_2^2}\right)^2}{2\frac{\sigma_1^2\sigma_2^2}{\sigma_1^2+\sigma_2^2}}-\frac{\left(\mu_1-\mu_2\right)^2}{2\sigma_1^2\sigma_2^2}\right)}\nonumber \\ 
        &= \frac{\exp{\left(-\frac{\left(\mu_{1}-\mu_{2}\right)^{2}}{2\left(\sigma_{1}^{2}+\sigma_{2}^{2}\right)}\right)}}{\sqrt{2 \pi\left(\sigma_{1}^{2}+\sigma_{2}^{2}\right)}} \cdot \frac{1}{\sqrt{2 \pi} \frac{\sigma_{1} \sigma_{2}}{\sqrt{\sigma_{1}^{2}+\sigma_{2}^{2}}}} \exp{\left(-\frac{\left(x-\frac{\mu_{1} \sigma_{2}^{2}+\mu_{2} \sigma_{1}^{2}}{\sigma_{1}^{2}+\sigma_{2}^{2}}\right)^{2}}{2 \frac{\sigma_{1}^{2} \sigma_{2}^{2}}{\sigma_{1}^{2}+\sigma_{2}^{2}}}\right)}\nonumber \\
        &= A\cdot \frac{1}{\sqrt{2 \pi} \frac{\sigma_{1} \sigma_{2}}{\sqrt{\sigma_{1}^{2}+\sigma_{2}^{2}}}} \exp\left(-\frac{\left(x-\frac{\mu_{1} \sigma_{2}^{2}+\mu_{2} \sigma_{1}^{2}}{\sigma_{1}^{2}+\sigma_{2}^{2}}\right)^{2}}{2 \frac{\sigma_{1}^{2} \sigma_{2}^{2}}{\sigma_{1}^{2}+\sigma_{2}^{2}}}\right). \label{eqn.p1}
    \end{align}
    Eqn.~\ref{eqn.p1} can be seen as PDF of $\mathcal{N}(\mu,\sigma)$  times $A$ where $\mu = (\frac{\mu_1}{\sigma_1^2}+\frac{\mu_2}{\sigma_2^2})\sigma^2,
    \frac{1}{\sigma^2} = \frac{1}{\sigma_1^2} + \frac{1}{\sigma_2^2}.$
    
    \item Induction step: Suppose it is true when $N=n$, and the product distribution of $n$ Gaussian experts has mean $\tilde{\mu}=(\frac{\mu_1}{\sigma_1^2}+\cdots+\frac{\mu_n}{\sigma_n^2})\tilde{\sigma}^2$ and variance $\frac{1}{\tilde{\sigma}^2}=\frac{1}{\sigma_1^2} + \cdots+\frac{1}{\sigma_n^2}$, then for $n+1$ Gaussian experts:
   \begin{equation}
\begin{aligned}
        \frac{1}{\sigma^2} &= \frac{1}{\tilde{\sigma}^2} + \frac{1}{\sigma_{n+1}^2} = \frac{1}{\sigma_1^2} + \cdots+\frac{1}{\sigma_n^2}+ \frac{1}{\sigma_{n+1}^2},\\
        \mu &= \Big(\frac{\tilde{\mu}}{\tilde{\sigma}^2}+\frac{\mu_{n+1}}{\sigma_{n+1}^2}\Big)\sigma^2 = \Big(\frac{\mu_1}{\sigma_1^2}+\cdots+\frac{\mu_n}{\sigma_n^2}+\frac{\mu_{n+1}}{\sigma_{n+1}^2}\Big)\sigma^2.
\end{aligned}
      \end{equation} 
    \item Eqn.~\ref{eqn.qwe} has been proved by the above derivation.
\end{itemize} 
\end{proof}

If we write $T_{ij}=(\sigma_{ij}^2)^{-1}$, then Eqn.~\ref{eqn.qwe} can be written as:
   \begin{equation}
\begin{aligned}
    \mu_i &= \Big(\sum_{j=1}^N \mu_{ij}T_{ij}\Big)\Big(\sum_{j=1}^N T_{ij}\Big)^{-1}, \\
    \sigma_i^2 &= \Big(\sum_{j=1}^N T_{ij}\Big)^{-1},
\end{aligned}
      \end{equation}

and is exactly what we're trying to prove.

\begin{algorithm}[h]
    \caption{CroMAC}
    \label{alg:AOS}
    \renewcommand{\algorithmicrequire}{\textbf{Input:}}
    \renewcommand{\algorithmicensure}{\textbf{Initialize:}}
    
    \begin{algorithmic}[1] \label{cCroMAC}
        \REQUIRE env, $\epsilon$, $\kappa$, $C\_MIN$, $C\_MAX$, $\alpha_1$, $\alpha_2$, $\alpha_3$, $T$,  $T_r$  
        \ENSURE Randomly initialize $\boldsymbol{\theta}$, $\bm \psi_{\rm enc}$, $\bm \psi_{\rm dec}$, $\bm \phi_{\rm enc}$, and initialize empty replay buffer $D$    
        
        \STATE  $t=0$
        
        \WHILE{$t<T$}
            \STATE Collect trajectory $h$ from env with $\boldsymbol{\theta}(\mathbf{a}|\phi_{\rm enc}(\mathbf{s}))$ and update replay buffer $D$
            \FOR{$j=1,2,\cdots$}
                \STATE Sample a batch of Episodes from $D$
                \STATE Update policy network:
                \STATE \quad $\boldsymbol{\theta} \leftarrow \boldsymbol{\theta} -\alpha_{\boldsymbol{\theta}} \nabla_{\boldsymbol{\theta}}\mathcal{L}(\boldsymbol{\theta})$ 
                \STATE Update VAE $\Phi$:
                \STATE \quad $\bm \psi_{\rm enc}\leftarrow \bm\psi_{\rm enc}-\alpha_{1}\nabla_{\bm\psi_{\rm enc}}\left(\mathcal{L}(\bm\psi)+\mathcal{L}(\boldsymbol{\theta})\right)$
                \STATE \quad $\bm\psi_{\rm dec}\leftarrow \bm\psi_{\rm dec}-\alpha_{1}\nabla_{\bm\psi_{\rm dec}}\mathcal{L}(\bm\psi)$
                \STATE Update MVAE $\phi$:
                \STATE \quad $\bm\phi_{\rm enc}\leftarrow \bm\phi_{\rm enc}-\alpha_{2}\nabla_{\bm\phi_{\rm enc}}\mathcal{L}_(\bm\phi)$
                \STATE \quad Clamp: $\bm\phi_{\rm enc}\leftarrow clamp(\bm\phi_{\rm enc},C\_MIN,C\_MAX)$
                \IF{$t>T_r$}
                    \STATE Update policy network:
                    \STATE \quad $\boldsymbol{\theta} \leftarrow \boldsymbol{\theta} -\alpha_{3} \nabla_{\boldsymbol{\theta}}\mathcal{L}_{\rm adv}$ 
                \ENDIF
                \STATE Update the target network $\boldsymbol{\theta}^-$ at regular intervals 
            \ENDFOR
            \STATE Update $t$
        \ENDWHILE
    \end{algorithmic}
\end{algorithm}

\section{Algorithm} \label{algorithm}
The whole optimization process is shown in Alg.~\ref{cCroMAC}. Where lines 6 to 7 are used to train policy network with only TD-error such that it has nothing to do with robust training; lines 8 to 10 aim at encoding the state into a latent space, while lines 11 to 13 train the MVAE with only partial observation and the received messages for decentralized execution, where $clamp(input,min,max)$ means clamping all elements in $input$ into range $[min,max]$; we train the policy network to be robust with auxiliary loss from lines 14 to 17. 

\section{Implemention Detail and Hyperparameters } 
\label{detailedimplement}
We train our CroMAC agents based on PYMARL with its default network structure and hyperparameters setting, except that different environments have different RNN hidden sizes. We use Adam optimizer with learning rate 0.0005 and other default hyperparameters. For the state encoder and decoder, we use an MLP with one
hidden layer and ReLU activation. For the message encoder, we use the same structure as the state encoder with shared parameters. All the prior distributions in the experiment are set to standard normal distribution, i.e., $\mathcal{N}(\mathbf{0},\mathbf{1})$. The other  hyperparameters of our proposed CroMAC for different benchmarks are summarized in Tab.~\ref{tab:hyperpara}.

Fast Gradient Sign Method (FGSM) ~\cite{goodfellow2014explaining} is a popular white-box method of generating adversarial examples, for MA-Dec-POMDP-Com, agent $i$ receives multiple messages $m_{ij}^t$ from teammate $j\in \{1, \cdots, i - 1, i + 1, \cdots, N\}$ at time $t$, we can compute perturbations for each $m_{ij}^t$ with individual Q-network $\theta_i$:
\begin{align*}
    \eta_{ij}^t = \boldsymbol{\epsilon} \cdot \text{sign} (\nabla_{m_{ij}^t}J(\theta_i,m_{ij}^t,y)),
\end{align*}
where $y$ is the original selected action and the perturbed example is:
$$\hat{m}_{ij}^t = m_{ij}^t+\eta_{ij}^t.$$
Projected Gradient Descent (PGD)~\cite{madry2017towards} can be regarded as an advanced version of FGSM where we implement it by adding perturbations within budget $\frac{\boldsymbol{\epsilon}}{3}$ to original message $3$ times.

\begin{table*}[h]
\caption{Hyperparameters in experiments.}
\centering
\begin{tabular}{c|c|c|c|c}
\midrule
\multirow{2}{*}{\begin{tabular}[c]{@{}c@{}}Hyperparameter \\ \end{tabular}} & \multirow{2}{*}{\begin{tabular}[c]{@{}c@{}}Hallway: 4x5x6 \\ Hallway: 3x3x4x4\end{tabular}} & \multirow{2}{*}{\begin{tabular}[c]{@{}c@{}}LBF: 3p-1f \\ LBF: 4p-1f\end{tabular}} & \multirow{2}{*}{\begin{tabular}[c]{@{}c@{}}TJ: slow \\ TJ: fast\end{tabular}} & \multirow{2}{*}{\begin{tabular}[c]{@{}c@{}}1o10b\_vs\_1r \\ 1o2r\_vs\_4r\end{tabular}} \\
& & & & \\
\midrule
\multirow{2}{*}{\begin{tabular}[c]{@{}c@{}}RNN  Hidden Dim \\ \end{tabular}} & \multirow{2}{*}{\begin{tabular}[c]{@{}c@{}}16 \\  \end{tabular}} & \multirow{2}{*}{\begin{tabular}[c]{@{}c@{}}32 \\  \end{tabular}} & \multirow{2}{*}{\begin{tabular}[c]{@{}c@{}}32 \\  \end{tabular}} & \multirow{2}{*}{\begin{tabular}[c]{@{}c@{}}64 \\  \end{tabular}} \\
& & & & \\
\midrule
\multirow{2}{*}{\begin{tabular}[c]{@{}c@{}}Z Dim \\ \end{tabular}} & \multirow{2}{*}{\begin{tabular}[c]{@{}c@{}}16 \\  \end{tabular}} & \multirow{2}{*}{\begin{tabular}[c]{@{}c@{}}32 \\  \end{tabular}} & \multirow{2}{*}{\begin{tabular}[c]{@{}c@{}}32 \\  \end{tabular}} & \multirow{2}{*}{\begin{tabular}[c]{@{}c@{}}64 \\  \end{tabular}} \\
& & & & \\
\midrule
\multirow{2}{*}{\begin{tabular}[c]{@{}c@{}}VAE Hidden  Dim \\ \end{tabular}} & \multirow{2}{*}{\begin{tabular}[c]{@{}c@{}}32 \\  \end{tabular}} & \multirow{2}{*}{\begin{tabular}[c]{@{}c@{}}64 \\  \end{tabular}} & \multirow{2}{*}{\begin{tabular}[c]{@{}c@{}}64 \\  \end{tabular}} & \multirow{2}{*}{\begin{tabular}[c]{@{}c@{}}128 \\  \end{tabular}} \\
& & & & \\
\midrule
\multirow{2}{*}{\begin{tabular}[c]{@{}c@{}}$\alpha_1$\\ \end{tabular}} & \multirow{2}{*}{\begin{tabular}[c]{@{}c@{}}0.1 \\  \end{tabular}} & \multirow{2}{*}{\begin{tabular}[c]{@{}c@{}}0.01 \\  \end{tabular}} & \multirow{2}{*}{\begin{tabular}[c]{@{}c@{}}0.01 \\  \end{tabular}} & \multirow{2}{*}{\begin{tabular}[c]{@{}c@{}}0.01 \\  \end{tabular}} \\
& & & & \\
\midrule
\multirow{2}{*}{\begin{tabular}[c]{@{}c@{}}$\alpha_2$\\ \end{tabular}} & \multirow{2}{*}{\begin{tabular}[c]{@{}c@{}}0.001 \\  \end{tabular}} & \multirow{2}{*}{\begin{tabular}[c]{@{}c@{}}0.001 \\  \end{tabular}} & \multirow{2}{*}{\begin{tabular}[c]{@{}c@{}}0.001 \\  \end{tabular}} & \multirow{2}{*}{\begin{tabular}[c]{@{}c@{}}0.01 \\  \end{tabular}} \\
& & & & \\
\midrule
\multirow{2}{*}{\begin{tabular}[c]{@{}c@{}}$\alpha_3$\\ \end{tabular}} & \multirow{2}{*}{\begin{tabular}[c]{@{}c@{}}0.3 \\ \end{tabular}} & \multirow{2}{*}{\begin{tabular}[c]{@{}c@{}}0.3 \\ \end{tabular}} & \multirow{2}{*}{\begin{tabular}[c]{@{}c@{}}0.3 \\ \end{tabular}} & \multirow{2}{*}{\begin{tabular}[c]{@{}c@{}}0.3 \\ 0.1\end{tabular}} \\
& & & & \\
\midrule
\multirow{2}{*}{\begin{tabular}[c]{@{}c@{}}$\kappa$\\ \end{tabular}} & \multirow{2}{*}{\begin{tabular}[c]{@{}c@{}}5 \\ 10\end{tabular}} & \multirow{2}{*}{\begin{tabular}[c]{@{}c@{}}5 \\ 10\end{tabular}} & \multirow{2}{*}{\begin{tabular}[c]{@{}c@{}}10 \\ \end{tabular}} & \multirow{2}{*}{\begin{tabular}[c]{@{}c@{}}10 \\ \end{tabular}} \\
& & & & \\
\midrule
\multirow{2}{*}{\begin{tabular}[c]{@{}c@{}}$C\_MAX$\\ \end{tabular}} & \multirow{2}{*}{\begin{tabular}[c]{@{}c@{}}0.1 \\ 0.2\end{tabular}} & \multirow{2}{*}{\begin{tabular}[c]{@{}c@{}}0.3 \\ \end{tabular}} & \multirow{2}{*}{\begin{tabular}[c]{@{}c@{}}0.3 \\ 0.6\end{tabular}} & \multirow{2}{*}{\begin{tabular}[c]{@{}c@{}}0.3 \\ 0.2\end{tabular}} \\
& & & & \\
\midrule
\multirow{2}{*}{\begin{tabular}[c]{@{}c@{}}Robust Start Time $T_r$\\ \end{tabular}} & \multirow{2}{*}{\begin{tabular}[c]{@{}c@{}}0.7M \\ \end{tabular}} & \multirow{2}{*}{\begin{tabular}[c]{@{}c@{}}0.8M \\ \end{tabular}} & \multirow{2}{*}{\begin{tabular}[c]{@{}c@{}}1.0M \\ \end{tabular}} & \multirow{2}{*}{\begin{tabular}[c]{@{}c@{}}1.0M \\ \end{tabular}} \\
& & & & \\
\midrule
\multirow{2}{*}{\begin{tabular}[c]{@{}c@{}}$\mathbb{\epsilon}$ of FGSM (1)\\ \end{tabular}} & \multirow{2}{*}{\begin{tabular}[c]{@{}c@{}}0.3 \\ $\setminus$\end{tabular}} & \multirow{2}{*}{\begin{tabular}[c]{@{}c@{}}0.02 \\ $\setminus$\end{tabular}} & \multirow{2}{*}{\begin{tabular}[c]{@{}c@{}}0.0003 \\ $\setminus$\end{tabular}} & \multirow{2}{*}{\begin{tabular}[c]{@{}c@{}}0.0055 \\ $\setminus$\end{tabular}} \\
& & & & \\
\midrule
\multirow{2}{*}{\begin{tabular}[c]{@{}c@{}}$\mathbb{\epsilon}$ of FGSM (2)\\ \end{tabular}} & \multirow{2}{*}{\begin{tabular}[c]{@{}c@{}}0.4 \\ $\setminus$\end{tabular}} & \multirow{2}{*}{\begin{tabular}[c]{@{}c@{}}0.25 \\ $\setminus$\end{tabular}} & \multirow{2}{*}{\begin{tabular}[c]{@{}c@{}}0.0004\\ $\setminus$\end{tabular}} & \multirow{2}{*}{\begin{tabular}[c]{@{}c@{}}0.0065 \\ $\setminus$\end{tabular}} \\
& & & & \\
\midrule
\multirow{2}{*}{\begin{tabular}[c]{@{}c@{}}$\mathbb{\epsilon}$ of FGSM \\ \end{tabular}} & \multirow{2}{*}{\begin{tabular}[c]{@{}c@{}}0.5 \\ \end{tabular}} & \multirow{2}{*}{\begin{tabular}[c]{@{}c@{}}0.03 \\ 0.05\end{tabular}} & \multirow{2}{*}{\begin{tabular}[c]{@{}c@{}}0.0005 \\ 0.001\end{tabular}} & \multirow{2}{*}{\begin{tabular}[c]{@{}c@{}}0.0075 \\ 0.015\end{tabular}} \\
& & & & \\
\midrule
\multirow{2}{*}{\begin{tabular}[c]{@{}c@{}}$\mathbb{\epsilon}$ of FGSM (3)\\ \end{tabular}} & \multirow{2}{*}{\begin{tabular}[c]{@{}c@{}}0.6 \\ $\setminus$\end{tabular}} & \multirow{2}{*}{\begin{tabular}[c]{@{}c@{}}0.35 \\ $\setminus$\end{tabular}} & \multirow{2}{*}{\begin{tabular}[c]{@{}c@{}}0.0006 \\ $\setminus$\end{tabular}} & \multirow{2}{*}{\begin{tabular}[c]{@{}c@{}}0.00875 \\ $\setminus$\end{tabular}} \\
& & & & \\
\midrule
\multirow{2}{*}{\begin{tabular}[c]{@{}c@{}}$\mathbb{\epsilon}$ of FGSM (4)\\ \end{tabular}} & \multirow{2}{*}{\begin{tabular}[c]{@{}c@{}}0.7 \\ $\setminus$\end{tabular}} & \multirow{2}{*}{\begin{tabular}[c]{@{}c@{}}0.4 \\ $\setminus$\end{tabular}} & \multirow{2}{*}{\begin{tabular}[c]{@{}c@{}}0.0007 \\ $\setminus$\end{tabular}} & \multirow{2}{*}{\begin{tabular}[c]{@{}c@{}}0.01 \\ $\setminus$\end{tabular}} \\
& & & & \\
\midrule
\end{tabular}
\label{tab:hyperpara}
\end{table*}

\section{Details about Baselines and Benchmarks}
\label{benchmarkdt}
We compare CroMAC against different baselines and variants on diverse multi-agent benchmarks, and we introduce more details about these baselines here.

\textbf{AME}.
Ablated Message Ensemble (AME) is a recently proposed certifiable defense, which can guarantee agents' robustness when only part of communication messages may suffer from perturbations. Specifically, AME makes a mild assumption that the attacker can only manipulate no more than half of the communication messages, and trains a message-ablation policy which takes in a subset of messages from other agents and outputs a base action. In deployment, an ensemble policy is introduced by aggregating multiple base actions from multiple subsets of messages which achieves the robustness goal to some extent. The pseudocode can be seen in Alg.\ref{aAME} and Alg.\ref{bAME}.

\begin{algorithm}[h]
    \caption{AME in the training phase}
    \label{alg:AOS}
    \renewcommand{\algorithmicrequire}{\textbf{Input:}}
    \renewcommand{\algorithmicensure}{\textbf{Initialize:}}
    
    \begin{algorithmic}[1] \label{aAME}
        \REQUIRE env, ablation size $k$  
        \ENSURE Randomly initialize $\hat{\pi}_i$ for each agent $i$.   
        \STATE  $t=0$
        
        \WHILE{$t<T$}
            \FOR{$i=1$ to $N$}
            \STATE Receive a set of messages $\rm \mathbf{m}_{:\rightarrow i}$, and update local history $\tau_i$
            \STATE Randomly sample a subset of messages $[\rm \mathbf{m}_{:\rightarrow i}]_k \sim Uniform (\mathcal{H}_k(\rm \mathbf{m}_{:\rightarrow i}))$
            \STATE Take action based on $\tau_i$ and the message subset $[\rm \mathbf{m}_{:\rightarrow i}]_k$, i.e., $a_i \leftarrow \hat{\pi}_i(\tau_i, [\rm \mathbf{m}_{:\rightarrow i}]_k)$
            \STATE Update replay buffer and policy $\hat{\pi}_i$
            \ENDFOR
        \ENDWHILE
    \end{algorithmic}
\end{algorithm}

\begin{algorithm}[h]
    \caption{AME in the testing phase}
    \label{alg:AOS}
    \renewcommand{\algorithmicrequire}{\textbf{Input:}}
    \renewcommand{\algorithmicensure}{\textbf{Initialize:}}
    
    \begin{algorithmic}[1] \label{bAME}
        \REQUIRE env, ablation size $k$,  trained message-ablation policy $\hat{\pi}_i$ for each agent $i$  
        \FOR{$i=1$ to $N$}
        \STATE Receive a set of messages $\rm \mathbf{m}_{:\rightarrow i}$ with no more than $C< \frac{N}{2}$ malicious messages, and update local history $\tau_i$
        \STATE Discrete action space:
        \STATE \quad Take action $a \leftarrow \arg \max_{a\in \mathcal{A}} \sum_{[\rm \mathbf{m}]_k \in \mathcal{H}_k(\mathbf{m})} \mathbb{I}[\hat{\pi}_i(\tau_i,[\mathbf{m}]_k)=a]$
        \STATE  Continuous Action Space:
        \STATE \quad Take action $a \leftarrow \rm Median \{\hat{\pi}_i(\tau_i,[\mathbf{m}]_k):[\mathbf{m}]_k \in \mathcal{H}_k(\mathbf{m}) \}$
        \ENDFOR
    \end{algorithmic}
\end{algorithm}

We introduce four types of testing environments as shown in Fig.~\ref{fig:envapp} in our paper, including Hallway~\cite{ndq}, Level-Based Foraging (LBF)~\cite{papoudakis2021benchmarking}, Traffic Junction (TJ)~\cite{tarmac}, and two maps named 1o2r\_vs\_4r and 1o10b\_vs\_1r requiring communication from StarCraft Multi-Agent Challenge (SMAC)~\cite{ndq}. In this part, we will describe the details of these used environments.

\textbf{Hallway}.
We design two instances of the Hallway environment, where agents can see nothing except its own location. In the first instance, we apply three hallways with lengths of $4, 5$, and $6$, respectively. That means we let three agents $a, b, c$ respectively initialized randomly at states $a_1$ to $a_4$, $b_1$ to $b_5$, and $c_1$ to $c_6$, and require them to arrive at state $g$ simultaneously. In the second instance, $4$ agents are distributed in four hallways with lengths of $3, 3, 4$, and $4$. A reward of $1$ will be given if all the agents arrives at the goal $g$ simultaneously. However, if any agent do not reach the 
goal $g$ at the same time, the game will stop immediately, and obtain $0$ reward.

\textbf{Level-Based Foraging (LBF)}.
We use a variant version of the original environment, where only one agent can can observe the map and others can see nothing. On this basis, we use two environment instances with different configurations, both of which are $8\times 8$ grid world, $1$ foods, and at least 3 agents are required to catch the food. One of them contains 3 agents and they need to complete the task in 25 time-steps while the other contains 4 agents but only 15 time-steps are provided for them.

\textbf{Traffic Junction (TJ)}.
We use the \textit{easy} version of the Traffic Junction environments as we mainly focus on the robustness of the algorithm. The \textit{slow} version has an agent number limit to $5$ and the max rate at which to add cars is $0.3$ while the \textit{fast} version has an agent number limit to $4$ and the max rate at which to add cars is $0.4$. In both of these two instances, the road dimension is $7$, and the sight of the agent is limited to $0$, which means each agent can only observe a $1\times 1$ field of view around it.

\textbf{StarCraft Multi-Agent Challenge (SMAC)}.
We use two maps named 1o2r\_vs\_4r and 1o10b\_vs\_1r in SMAC, which are introduced in NDQ~\cite{ndq}. In 1o2r\_vs\_4r, an Overseer finds $4$ Reapers, and the ally units, $2$ Roaches, need to reach enemies and kill them. Similarly, 1o10b\_vs\_1r is a map full of cliffs, where an Overseer detects a Roach, and the randomly spawned ally units, $10$ Banelings, are required to reach and kill the enemy.





\end{appendix}

\end{document}